%% file: wake.tex
\documentclass{article}
\usepackage[utf8]{inputenc}

\usepackage{graphicx}
\usepackage{amsmath}
\usepackage{amsfonts}
\usepackage{bm}
\usepackage{physics}

\usepackage{algorithm}
\usepackage{algpseudocode}

\algblock{Input}{EndInput}
\algnotext{EndInput}
\algblock{Output}{EndOutput}
\algnotext{EndOutput}
\newcommand{\Desc}[2]{\State \makebox[10em][l]{#1}#2}

\usepackage[version=4]{mhchem}
\usepackage{siunitx}
\usepackage{longtable,tabularx}
\usepackage{soul,color,mathtools}
\usepackage{caption}
\usepackage{subcaption}

\usepackage{fancyhdr}

\usepackage[
backend=biber,
style=numeric-comp,
sorting=ynt
]{biblatex}
\AtEveryBibitem{
    \clearfield{urlyear}
    \clearfield{urlmonth}
    \clearfield{urldate}
}

\fancypagestyle{alim}{\fancyhf{}\fancyfoot[C]{Distribution Statement A: Approved for Public Release; Distribution is Unlimited. PA\# AFRL-2024-0012.}}

\usepackage{subfiles} 
\begin{document}


\title{Acceleration of RANS Solver Convergence via Initialization with Wake Extension Models}

\author{Kazuko W. Fuchi$^1$, Eric M. Wolf$^2$, Christopher R. Schrock$^3$, and Philip S. Beran$^3$ }
\date{%
    $^1$University of Dayton Research Institute\\%
    $^2$Ohio Aerospace Institute/Parallax Advanced Research\\%
    $^3$Air Force Research Laboratory\\[2ex]%
}

\maketitle

\begin{abstract}

\subfile{sections/abstract}

\end{abstract}

\thispagestyle{alim}


\section{Introduction}\label{sec:intro}

\subfile{sections/introduction}

\section{Method}\label{sec:method}

\subfile{sections/method}

\section{Results}\label{sec:results}
\subfile{sections/results}

\section{Conclusion}\label{sec:conclusion}

\subfile{sections/conclusion}

\section*{Acknowledgments}
This work was sponsored by the Air Force Office of Scientific Research Computational Mathematics Program (Dr. Fariba Fahroo, Program Officer).

\printbibliography 

\section*{Appendix}

\subfile{sections/appendix}

\end{document}

%% file: sections/abstract.tex
Use of appropriate initialization to warm-start Reynolds-averaged Navier-Stokes (RANS) simulations of turbulent flow can facilitate convergence and lead to efficient use of computational resources. In this work, a method to model downstream wake development in external turbulent flow is proposed and used for RANS solver convergence acceleration. To balance the model accuracy and cost, the proposed method divides the analysis domain into three regions: near-body, wake and off-body. An approach based on a convolutional neural network is introduced as an efficient method to predict the downstream wake development. The model training only requires data from a single simulation, and its use is demonstrated to be effective in accelerating the RANS simulation when combined with an accurate flow prediction in the near-body region. The simulation using the proposed method took 26.3x fewer iterations, achieving 16.4x speedup in wall-clock time, compared to a baseline run using freestream initialization.

%% file: sections/introduction.tex
Steady-state solutions to Reynolds-averaged Navier-Stokes (RANS) simulations have an important role in performance evaluation during aerospace design~\cite{Spalart2016,Skinner2018}. However, RANS simulations are computationally expensive, and their integration in an iterative design process could prolong the design evaluation step significantly, potentially causing the overall computational time of design iterations to become prohibitively long. The considerable computational expense of performing RANS simulations has led to the application of machine learning (ML) techniques to augment~\cite{BelbutePeres2020} or even replace~\cite{Jin2021} traditional computational fluid dynamics (CFD) solvers, leading to models that can be evaluated more rapidly. However, physics-informed approaches~\cite{Cai2021,Eivazi2022a} have not yet proven to be more accurate or efficient than traditional solvers~\cite{Chuang2022,Grossmann2023}, while data-driven approaches may require a large amount of training data, needing perhaps tens, hundreds, or even thousands of CFD simulations to be performed to acquire the necessary training data~\cite{Guo2016,Afshar2019a,Thuerey2020,Kashefi2021,Chen2023d}. Furthermore, while trusted CFD solvers may have undergone extensive verification and validation campaigns~\cite{Oberkampf2010} and experienced users may use well-established best practices~\cite{Mendenhall2003} to obtain reliable results from CFD simulations within certain domains, the predictions of ML models may not be sufficiently accurate for scientific or engineering purposes, especially in cases that require extrapolation from the training data set~\cite{Tangsali2020}.

In light of this situation, a body of work has emerged in which ML model predictions are applied as initial conditions for traditional CFD solvers in order to accelerate the convergence of the solver, which we will refer to as the CFD initialization task. In this way, the trusted CFD solution is sought to be obtained at a reduced computational cost. While the CFD community has long recognized that the choice of initial condition has impacts on solver convergence and stability~\cite{Anderson1995} and even the qualitative nature of numerical solutions obtained at finite residuals~\cite{OllivierGooch2023}, the viability of a proposed ML-based approach to the CFD initialization task relies upon, first, how much acceleration of CFD simulations can be achieved, which determines the benefit, and, second, how well efforts required for data acquisition and model training can be managed, which determines the cost.

Some works have applied convolutional neural networks (CNNs) to predict flow fields on a deforming logically Cartesian structured mesh for the CFD initialization task~\cite{ObiolsSales2020,Tsunoda2022,Zuo2023}, reporting solver convergence speedups of approximately 2-4x for turbulent airfoil cases. Related works explored techniques for reducing the costs of such models. In~\cite{ObiolsSales2021}, a transfer learning approach was applied to reduce the cost of training data acquisition and model training of a CNN-based model. In~\cite{Tsunoda2023}, a heterogeneous computing architecture was applied in a multi-objective design optimization problem, with a CNN model being trained on GPU simultaneously with a CFD solver running on CPU, reducing overall wall-clock computation time to 57\% of the baseline optimization run.

These CNN-based approaches attempt to learn the flow field across an entire mesh, learning both important near-body boundary layer and wake regions as well as off-body regions that are close to freestream conditions. Other works have developed approaches that focus model learning on important flow features by restricting the model to a rectangular near-body region. In~\cite{Fuchi2022}, a neural network was trained to predict velocity and pressure fields for laminar flow around an airfoil in a near-body region using elliptic input features derived from potential flow solutions, with the predicted flow fields being used for the CFD initialization task, achieving approximately 2x speedup over freestream initialization. In~\cite{Zhou2024}, a vector cloud neural network with equivariance was trained to predict RANS solution fields in a near-body region based on potential flow input features, achieving an approximate 2x speedup over freestream initialization in the CFD initialization task. In both of these works, the model is extended outside of the near-body region with freestream conditions.

In this work, we investigate the effect of the size of the near-body region on the CFD initialization task for a case of turbulent flow around an airfoil. It is found that the number of iterations required for solver convergence decreases with the downstream extent of the near-body region, corresponding to the amount of the wake represented, with maximal speedup only being achieved when the entire wake is represented to the downstream mesh boundary. We investigate a number of wake extender models, which extend the predicted field of the near-body region in the downstream direction in order to model the wake to the downstream mesh boundary. We propose a CNN-based wake extender model that maps a flow field profile to a corresponding flow field profile at a further distance downstream. This CNN-based model is inexpensive to train, can be trained with data from a single CFD simulation, and gives excellent performance in the CFD initialization task, achieving a 26.3x reduction in time step count and a 16.4x speedup in wall-clock time over freestream initialization.

The remainder of the paper proceeds as follows. The methods used to train the wake model and initialize the solver are described in Sec. \ref{sec:method}. Demonstration of the proposed method for a turbulent flow around NACA0012 airfoil is provided in Sec. \ref{sec:results}, along with a comparison between our proposed CNN-based wake model and other wake models. The paper concludes with remarks on our findings in Sec. \ref{sec:conclusion}.

%% file: sections/method.tex
\begin{figure}[ht!]
\centering
    \begin{subfigure}{3in}
    \includegraphics[width=\textwidth]{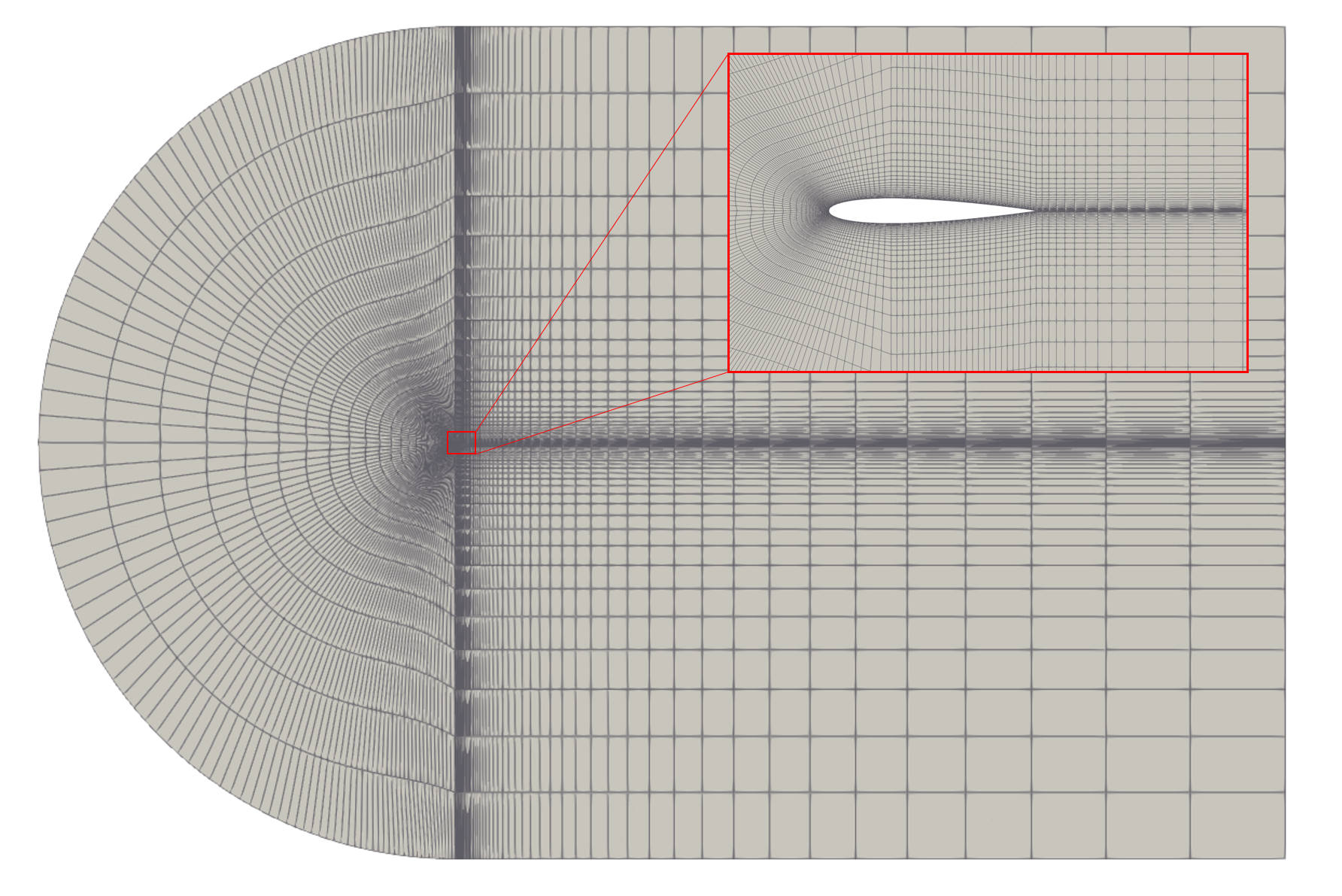}
        \caption{Mesh in the full domain}
    \end{subfigure}
    \begin{subfigure}{2.3in}
		\includegraphics[width=\textwidth]{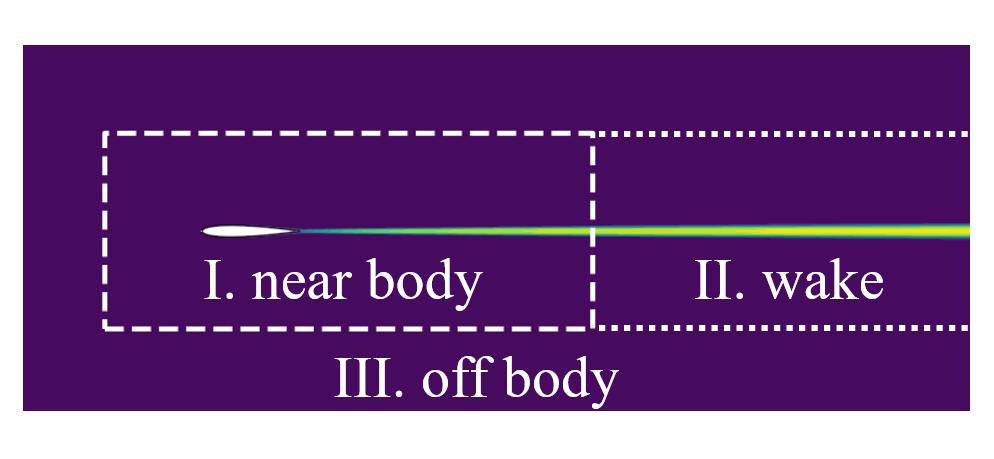}
        \caption{Subdomains}
    \end{subfigure}
\caption{Three modeling regions of the analysis domain.}\label{fig:regions}
\end{figure}

The CFD analysis domain for an external fluid flow typically extends far beyond the region immediately surrounding the body, to mitigate any spurious effect of the farfield boundary, as shown in Fig. \ref{fig:regions} (a). In this work, the domain is decomposed into three regions: I) near-body, II) wake and III) off-body regions, as illustrated in Fig. \ref{fig:regions} (b). 
While many existing works concentrate on the construction of near-body models applicable in region I, the current work focuses on the treatment of the wake in region II. The method proposed here is agnostic to the choice of model in the near-body region, and in order to establish an upper bound on performance in the CFD initialization task with respect to near-body model quality, interpolated high-fidelity (HF) solution data is used as a surrogate near-body model in the numerical experiments presented in this work. This isolates the investigation of the efficacy of wake modeling in the CFD initialization task from being impacted by the accuracy of the near-body model.

\subsection{Model Ensemble through POFU}\label{sec:method-pofu}

Flow fields in the three regions are assumed to be predicted from separate models and combined using the partition of unity functions (POFU) to prepare full-domain fields to generate CFD initialization data. For the purposes of this work, we define POFU functions as sets of functions over $\mathbb{R}^{2}$ which take values between $0$ and $1$ and sum to unity everywhere in $\mathbb{R}^{2}$.  

\begin{figure}[ht!]
    \centering
    \begin{subfigure}{2.4in}
        \includegraphics[width=\textwidth]{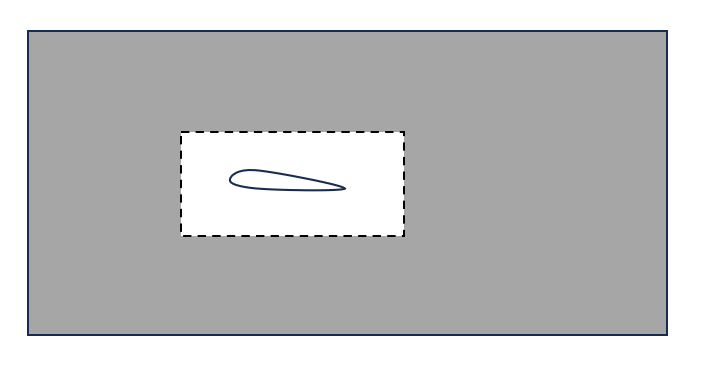}
        \caption{Near-body window function $W_{N}$}
    \end{subfigure}
    \begin{subfigure}{2.4in}
        \includegraphics[width=\textwidth]{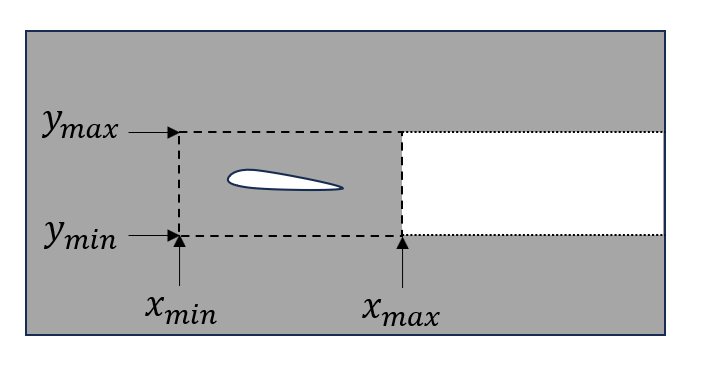}
        \caption{Wake window function $W_{W}$}
    \end{subfigure}
    \caption{POFU window functions for (a) near-body, and (b) wake regions, with $W=1$ in white and $W=0$ in gray areas.}\label{fig:pofu-window-funcs}
\end{figure}

The full flow fields $\tilde{\bm{q}}$ can be evaluated as an ensemble of near-body region, wake region, and off-body region models according to the following POFU construction:

\begin{equation}\label{eq:pofu_extension}
    {\tilde{\bm{q}}} = W_{N} \tilde{\bm{q}_{N}}+\left(1-W_{N}\right)\left[W_{W}\tilde{\bm{q}_{W}}+\left(1-W_{W}\right)\tilde{\bm{q}_{O}}\right], 
\end{equation}

\noindent
where $\tilde{\bm{q}_{N}}$, $\tilde{\bm{q}_{W}}$, and $\tilde{\bm{q}_{O}}$ refer to the predicted fields in the near-body, wake, and off-body regions, respectively. The POFU construction is based on the use of window functions to select desired models, as illustrated in Fig.~\ref{fig:pofu-window-funcs}. The near-body window function (e.g., Fig. \ref{fig:pofu-window-funcs} (a)) is used to select the near-body model inside region I, and the wake window function (e.g., Fig. \ref{fig:pofu-window-funcs} (b)) is used to select the wake model in region II. The POFU formulation enforces the off-body model in region III (the rest of the analysis domain).


\subsection{Wake Extension Models}\label{sec:method-wake}
\begin{figure}[ht!]
    \centering
    \begin{subfigure}{2.6in}
        \includegraphics[width=\textwidth]{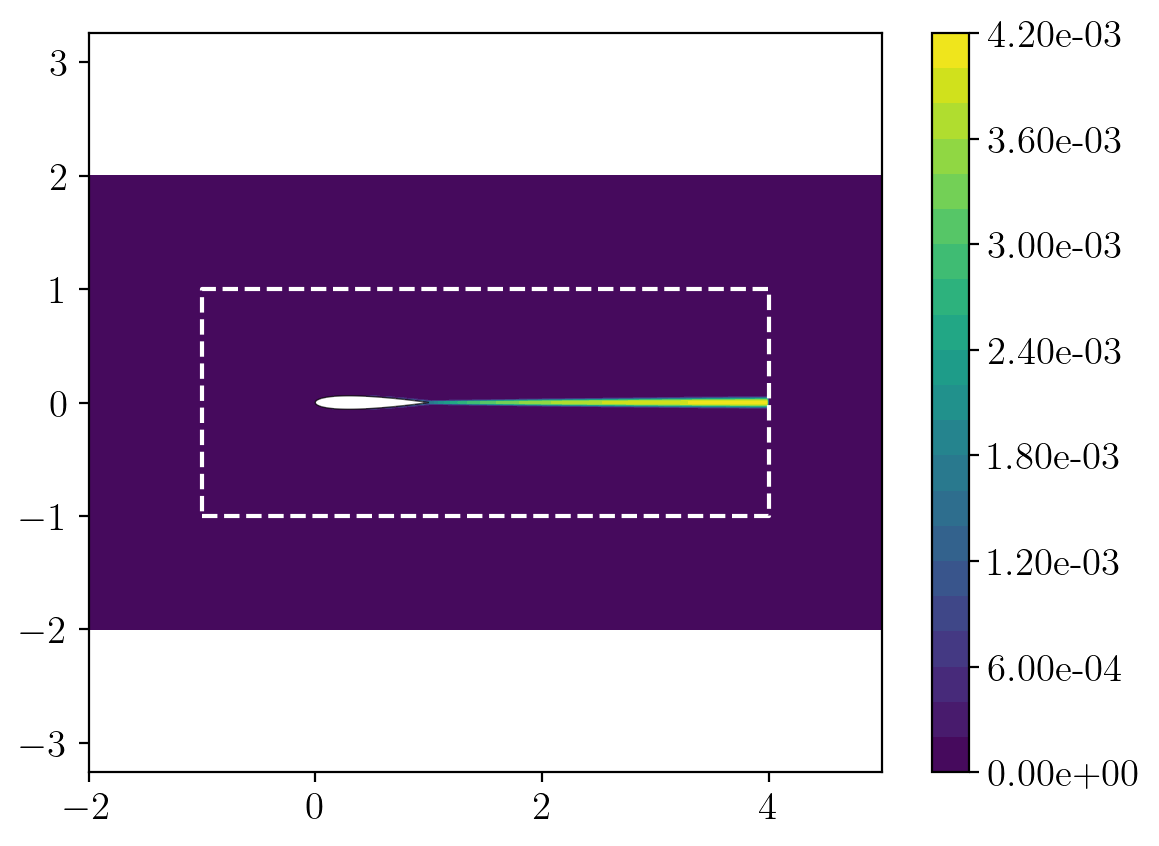}
        \caption{Freestream conditions}
    \end{subfigure}
    \begin{subfigure}{2.6in}
        \includegraphics[width=\textwidth]{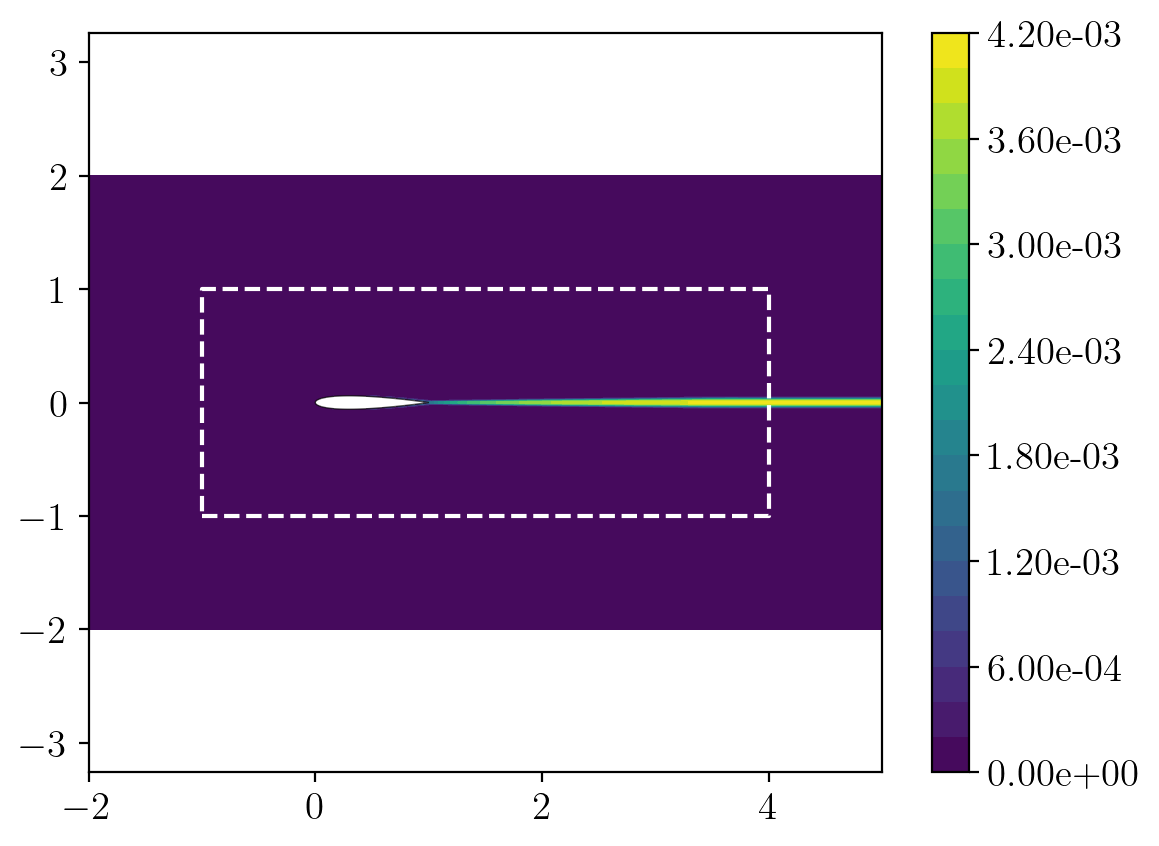}
        \caption{Uniform wake extension}
    \end{subfigure}
		\\
    \begin{subfigure}{2.6in}
        \includegraphics[width=\textwidth]{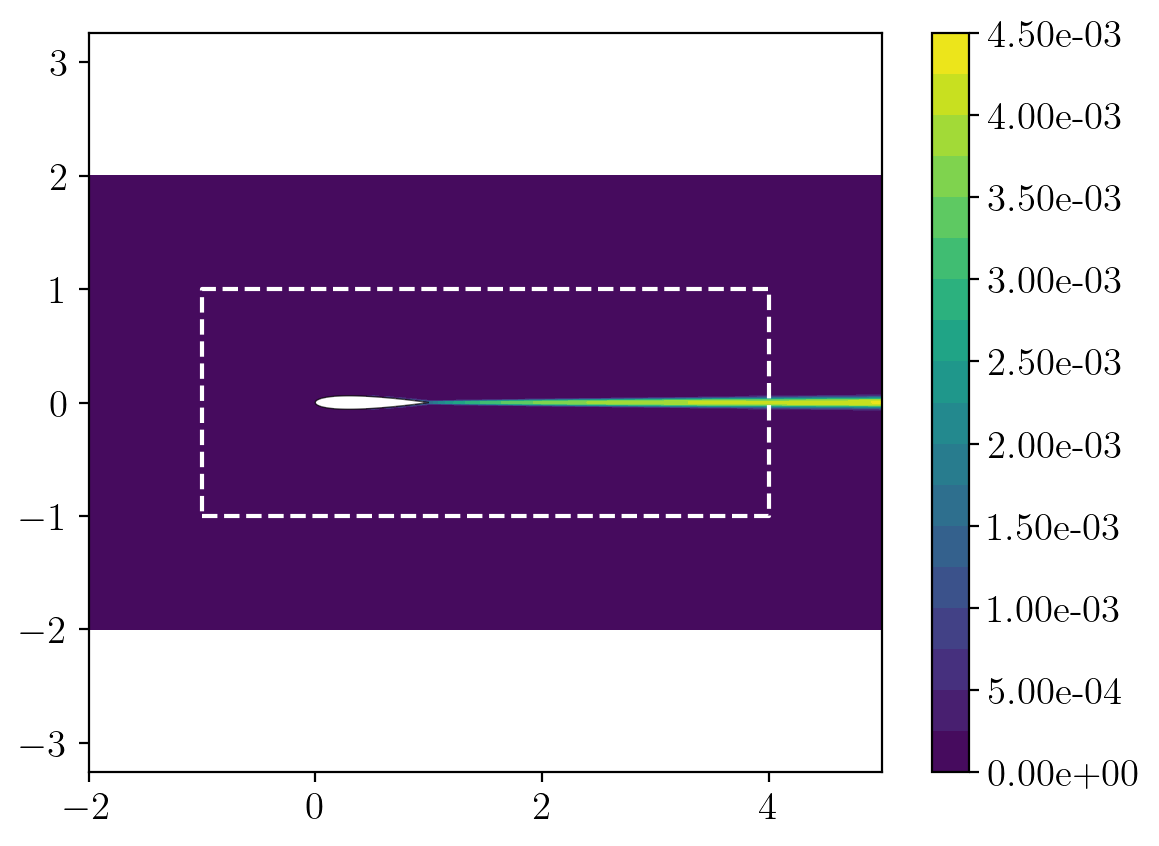}
        \caption{CNN wake extension}
    \end{subfigure}
    \caption{Examples of flow field using different wake models: (a) freestream conditions, (b) uniform wake extension, and (c) CNN wake extension model.}\label{fig:pofu-w-wake-extender}
\end{figure}

Two simple strategies for wake modeling: freestream conditions and uniform extension are described first, and a method based on a CNN is introduced. Examples of flow fields using the three wake model are shown in Fig. \ref{fig:pofu-w-wake-extender}.

\subsubsection{Freestream in the wake}
A simple strategy to handle the wake region without a specifically designed model is to use freestream conditions. In this model, the wake is cut off at the right end of the near-body region, and the flow field quantities are extended with freestream conditions outside of the near-body region, as seen in Fig. \ref{fig:pofu-w-wake-extender} (a). The size of the near-body region, or the location of the transition interface between near-body and wake regions has an impact on the CFD solver convergence time and will be discussed in Sec. \ref{sec:results-init}. 

\subsubsection{Uniform wake extension}
Another strategy is to uniformly extend the wake downstream based on a profile derived from the near-body window. The flow features along the interface between the near-body and wake regions from the near-body model can be used to generate uniformly extended wake. To reduce any local noise from the near-body model, the fields predicted at several $x-$slices within a small range can be averaged and used in this wake model.
In this model, the location of the interface between near-body and wake regions impacts the CFD initialization not only through the size of the near-body region but also the quality of the wake model as generated by the flow field profile at the selected interface location. 



\subsubsection{CNN wake extender model}
Our proposed model predicts downstream development of the wake using a CNN. The CNN architecture is particularly suitable for modeling wake development, as all modeling occurs away from the body and no special geometric accommodations are required, in contrast to the situation with near-body modeling, where body geometry must be encoded through mechanisms such as binary masking functions~\cite{Thuerey2020}, signed distance functions~\cite{Tangsali2020}, or mesh deformation~\cite{ObiolsSales2020,Tsunoda2022}.

\begin{figure}[ht!]
\centering
\includegraphics[width=2.8in]{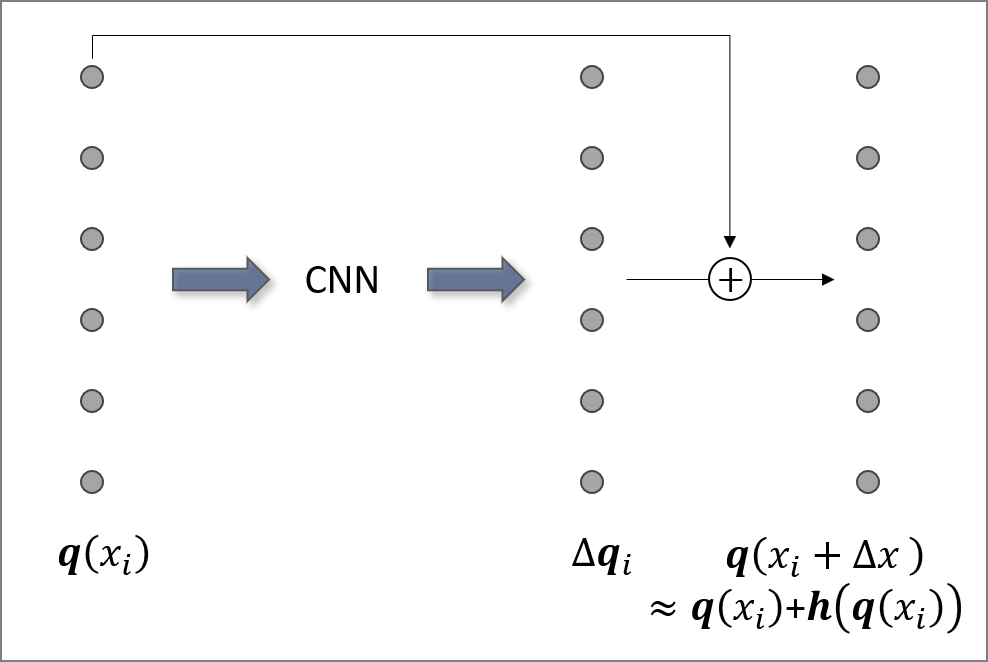}
\caption{CNN model is used to predict the field difference downstream.}\label{fig:cnn-schematic}
\end{figure}

\begin{figure}[ht!]
\centering
\includegraphics[width=3.2in]{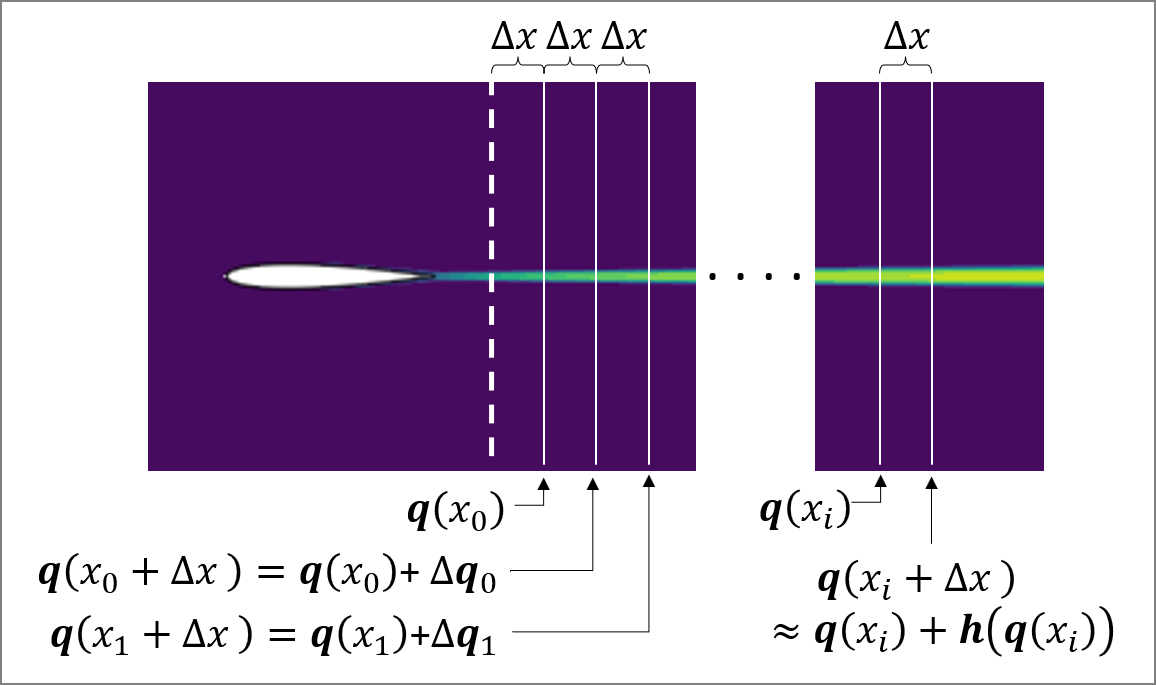}
\caption{Wake extension through the recursive use of CNN model.}\label{fig:cnn-data-prep}
\end{figure}

The CNN model is constructed such that the model accepts the flow field distribution along the interface of near-body and wake regions as input and predicts the wake development downstream. A simple schematic of the method is provided in Fig. \ref{fig:cnn-schematic}. 
The trained model accepts the field quantity $\bm{q}\left(x_{i}\right)$ at a set of $n$ points equidistributed along the slice $x=x_{i}$ and predicts the change in the field $\Delta \bm{q}_{i}=\bm{q}\left(x_{i+1}\right)-\bm{q}\left(x_{i}\right)$ for corresponding points at a prescribed distance $\Delta x$ downstream. The estimated change $\bm{h}_{i}\approx \Delta \bm{q}_{i}$ is then used to estimate the field quantity at $x_{i}+\Delta x$, as 

\begin{equation}\label{eq:cnn-wake}
    \bm{q}\left(x_{i}+\Delta x \right) = \bm{q}\left(x_{i}\right) + \Delta \bm{q}_{i} \approx \bm{q}\left(x_{i}\right) + \bm{h}_{i},
\end{equation}

\noindent
where $\bm{h}_{i}=\bm{h}\left(\bm{q}\left(x_{i}\right)\right)$ refers to the CNN model that approximates $\Delta \bm{q}_{i}$ as a function of $\bm{q}\left(x_{i}\right)$. The model is used recursively to predict the wake development from the interface to the end of the analysis domain, as illustrated in Fig. \ref{fig:cnn-data-prep}.

\begin{algorithm}
\captionof{algorithm}{CNN wake extension model training and usage}\label{alg:cnn-wake}
\begin{algorithmic}[1]

\Input
    \Desc{$\bm{q}\left(x_{0}\right)$}{flow fields at the interface $x_{0}$}
\EndInput
\Output
    \Desc{$\{\tilde{\bm{q}}\left(x_{i},y_{i}\right)\}_{i=1}^{N}$}{predicted flow fields at mesh center coordinates $\{\left(x_{i},y_{i}\right)\}_{i=1}^{N}$}
\EndOutput
\State Prepare flow fields $\bm{q}\left(x_{0}\right)$ along the interface $x_{0}$ via model evaluation, interpolation, or otherwise
\For{$i=0,\cdots , m-1$}
\State evaluate CNN model to approximate the change in the field $\Delta \bm{q}\left(x_{i}\right) \approx \bm{h}\left(\bm{q}\left(x_{i}\right)\right)$
\State extend the wake prediction by $\Delta x$ through $\bm{q}\left(x_{i+1}\right) = \bm{q}\left(x_{i}\right) +\Delta \bm{q}\left(x_{i}\right) \approx \bm{q}\left(x_{i}\right) +\bm{h}\left(\bm{q}\left(x_{i}\right)\right)$
\EndFor

\State Construct interpolator $\tilde{\bm{q}}\left(\bm{x},\bm{y}\right)$
\For{$i=1,\cdots , N$}
	\State evaluate predicted field $\tilde{\bm{q}}\left(x_{i},y_{i}\right)$ at mesh center coordinate $\left(x_{i},y_{i}\right)$ 
\EndFor
\end{algorithmic}
\end{algorithm}

%
%

The loss function for the model training is defined as:
\begin{equation}\label{eq:cnn-loss}
    \mathcal{L} = \frac{1}{m}\sum_{i=0}^{m-1}{\norm{\Delta \bm{q}_{i} - \bm{h}\left(\bm{q}\left(x_{i}\right)\right)}^{2}},
\end{equation}

\noindent
where $m$ denotes the number of $x-$slices, or data size, used in the model training.
The CNN training data is generated using $\Delta \bm{q}_{0}, \Delta \bm{q}_{1}, \cdots, \Delta \bm{q}_{m-1}$  from the HF solution. 

\begin{figure}[ht!]
\centering
\includegraphics[width=3.4in]{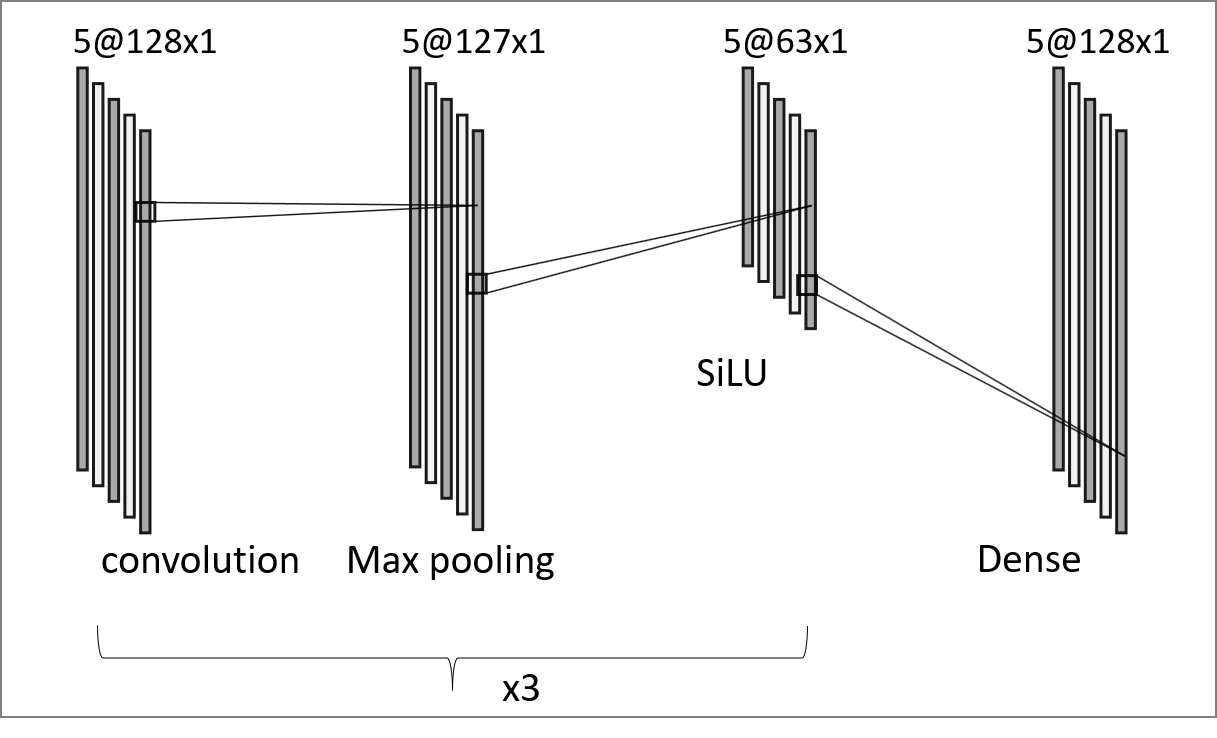}
\caption{CNN architecture used for the wake extension model.}\label{fig:cnn-architecture}
\end{figure}

A schematic of the CNN architecture used in this work is prepared using a web tool by LeNail~\cite{LeNail2019} and shown in Fig. \ref{fig:cnn-architecture}. The input data consist of field values at points along the slices $x=x_{0}, x_{1}, \cdots, x_{m-1}$. Along each of the $m$ slices, the field is evaluated at $n$ points between the specified lower and upper limits of the wake region, $\left[ y_{min},y_{max} \right]$. The output data are the corresponding field differences $\Delta \bm{q}_{0},\Delta \bm{q}_{1},\cdots,\Delta \bm{q}_{m-1}$. Data size is determined by the number of $x-$slices $m$ used to generate data.
In Fig. \ref{fig:cnn-architecture}, a convolutional layer is applied to the input layer, followed by a max pooling layer and an activation layer. The convolution-max pooling-activation layers are repeated multiple times, and a linear output layer is applied at the end. The example in Fig. \ref{fig:cnn-architecture} represents a case with $n=128$ evaluation points, five channels corresponding to five flow field quantities, and three convolution-activation-max pooling repetitions.

\begin{figure}[ht!]
\centering
\includegraphics[width=4.in]{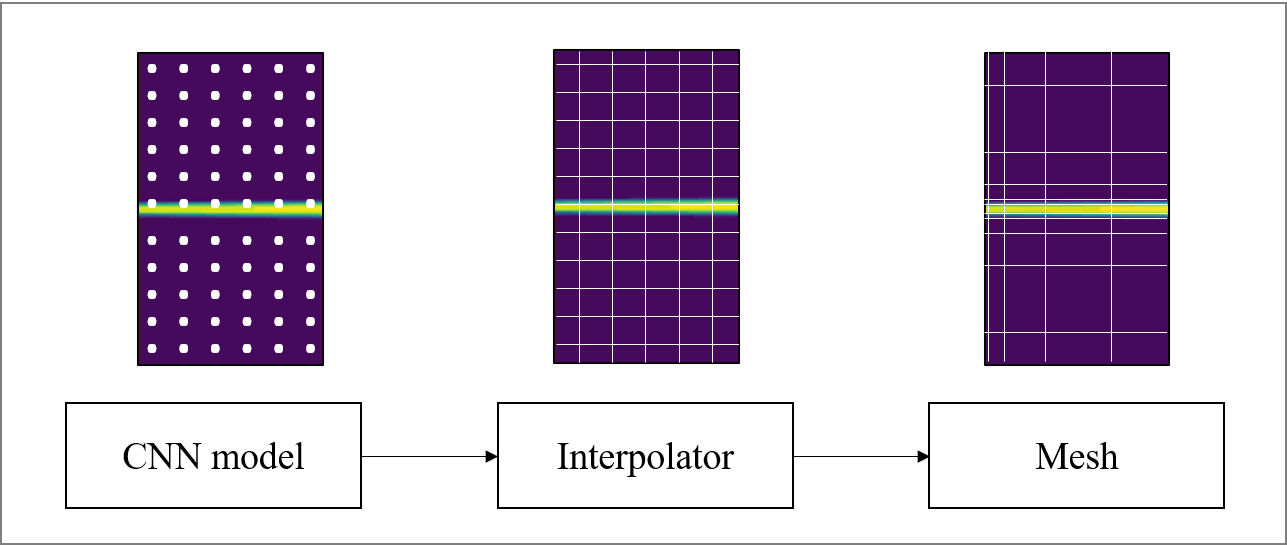}
\caption{CNN wake model output is interpolated onto CFD mesh.}\label{fig:cnn-cfd-init}
\end{figure}

Due to the model design, flow field quantities in the wake region are predicted on a uniform Cartesian grid. The resolution of the grid is determined by the number of points $n$ along the interface between the near-body and wake regions and the inter-slice distance $\Delta x$. A linear interpolator is constructed based on this output and used to estimate the fields over the CFD mesh, as illustrated in Fig. \ref{fig:cnn-cfd-init}.

%

The overall algorithm for training the CNN wake extension model and using the trained model to prepare the CFD initialization data is provided in Algorithm \ref{alg:cnn-wake}.

\subsection{ML Initialization of CFD Simulations}\label{sec:method-cfd}
\begin{figure}[ht!]
\centering
\includegraphics[width=4.6in]{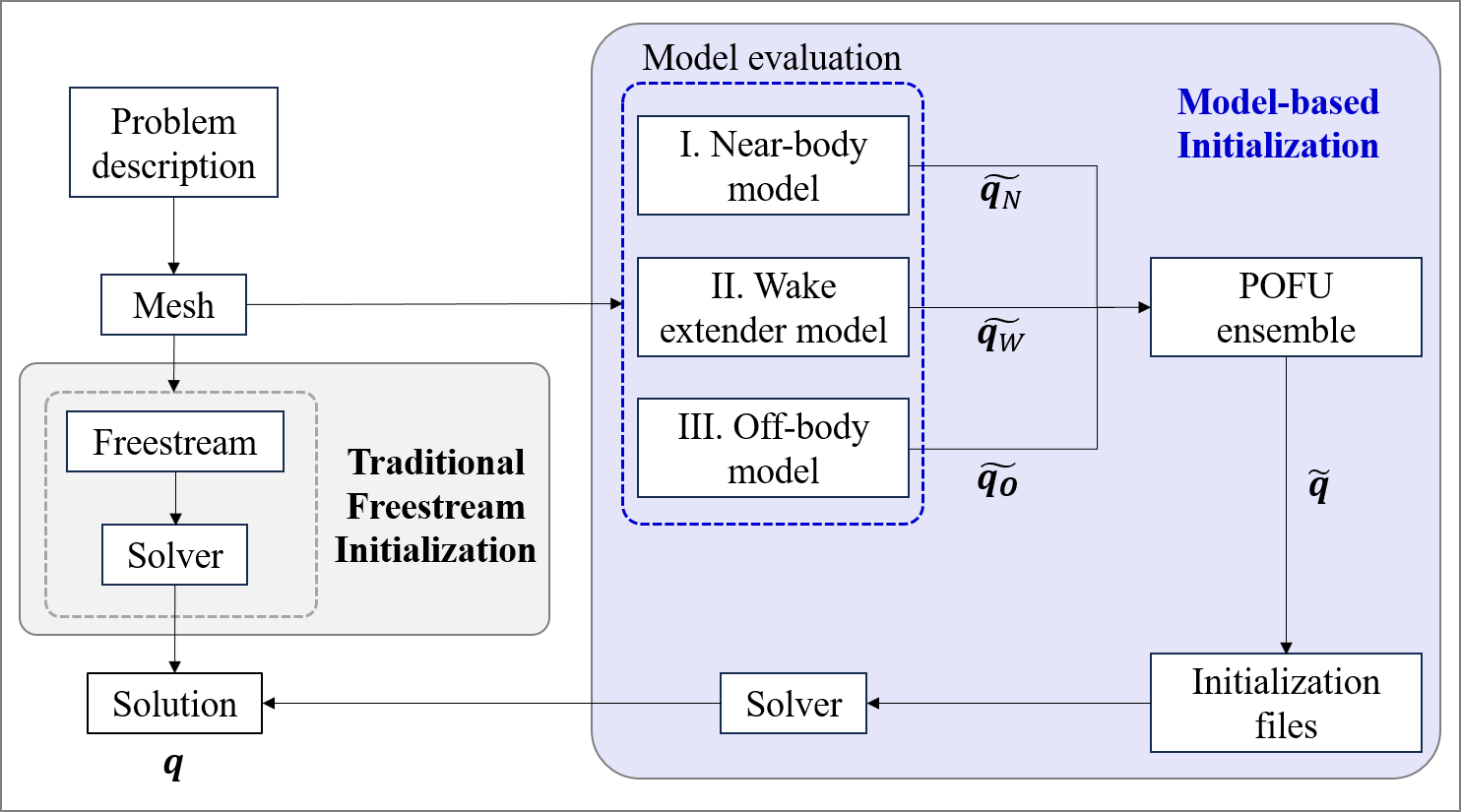}
\caption{Flow of operations for model-based CFD initialization.}\label{fig:flowchart-cfd-init}
\end{figure}

A general outlook of the process to carry out the CFD initialization task is shown in Fig. \ref{fig:flowchart-cfd-init}. 
First, a CFD mesh appropriately selected for the target flow problem is prepared.
The near-body and wake models are evaluated to estimate the flow fields in the near-body and wake regions, respectively. 
%
The predicted fields in the two regions, along with the fields from the off-body model (freestream conditions in this work), are combined through the POFU formulation in Eq. (\ref{eq:pofu_extension}) and evaluated on the mesh. The data are then written into restart or initialization files that can be read by the selected CFD solver, and the simulation is performed. Depending on the solver implementation, additional processes such as conversion of grid point to cell values or computation of secondary or derived quantities may be necessary. Further details of the process used in RANS simulations of flow around a NACA0102 airfoil in OpenFOAM are described in Sec. \ref{sec:results}.

%% file: sections/results.tex

This section describes the data preparation and the CNN wake extension model training, followed by the analysis of CFD initialization results obtained using the three wake extension models described in Sec. \ref{sec:method-wake} \footnotemark.

\footnotetext{
    All computations were performed on a dual socket Intel(R) Xeon(R) CPU E5-2637 v4 workstation with 8 total cores and 96GB of RAM.}

\subsection{Data Preparation and Wake Model Training}


\subsubsection{Data preparation - turbulent flow around NACA0012 airfoil}\label{sec:results-naca0012}

\begin{figure}[htb!]
    \centering
    \begin{subfigure}{2.6in}
        \includegraphics[width=\textwidth]{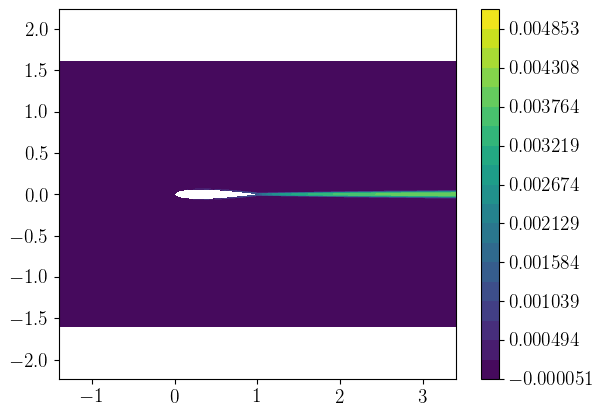}
        \caption{$\tilde{\nu}$}
    \end{subfigure}
		\begin{subfigure}{2.6in}
    \includegraphics[width=\textwidth]{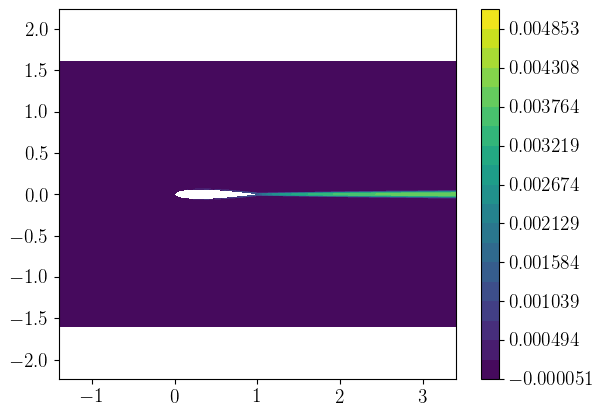}
    \caption{$\nu_{t}$}
    \end{subfigure}
		\\
    \begin{subfigure}{2.6in}
        \includegraphics[width=\textwidth]{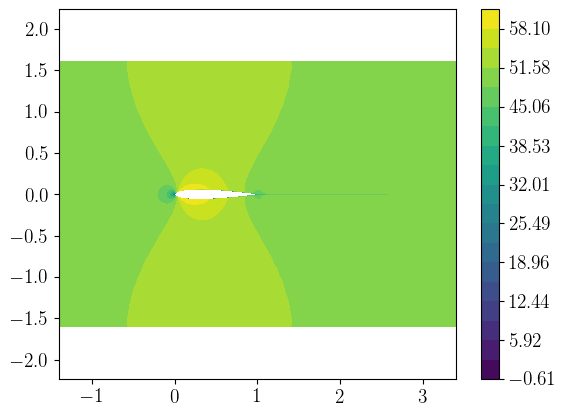}
        \caption{$u$}
    \end{subfigure}
    \begin{subfigure}{2.6in}
        \includegraphics[width=\textwidth]{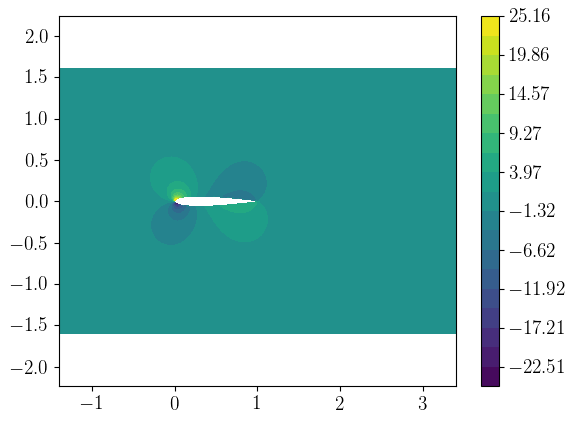}
        \caption{$v$}
    \end{subfigure}
		\\
		\begin{subfigure}{2.6in}
    \includegraphics[width=\textwidth]{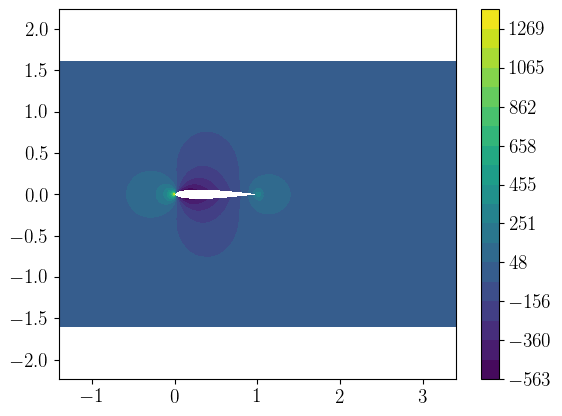}
    \caption{$p$}
    \end{subfigure}
    \caption{HF solution for incompressible turbulent flow around NACA0012 airfoil at $Re=6\times 1e^{6}$ and $M=$0.15.}\label{fig:hf-soln}
\end{figure}

Turbulent flow around a NACA0012 airfoil is simulated in OpenFOAM and used to demonstrate the proposed method. We consider the 2D NACA 0012 Airfoil Validation Case as defined by the NASA Turbulence Modeling Resource website~\cite{nasatmr_naca0012case}, which has Reynolds number $Re=6\times 1e^{6}$ and Mach number $M=0.15$, while the chord length is set to 1. As this case is essentially incompressible, OpenFOAMs \texttt{simpleFoam}, a steady-state solver for incompressible, turbulent flow~\cite{noauthor_openfoam_nodate} is used  with the solver and numerical scheme settings shown in Tab. \ref{tab:openfoam-setting} to solve the incompressible RANS equations with the Spalart-Allmaras turbulence model.
Data from five computed flow field quantities, $x-$component of the velocity $u$, $y-$component of the velocity $v$, pressure $p$, Spalart-Allmaras modified viscosity $\tilde{\nu}$ and turbulent kinematic viscosity $\nu_{t}$,  are computed and collected. 
The mesh consists of 16200 cells, and inlet and outlet boundaries are placed 50x and 100x chord lengths away from the body, respectively.
Flow fields of the converged solution, obtained through the standard solution process using freestream conditions for the initialization, are shown in Fig. \ref{fig:hf-soln}. The solution is saved into \texttt{vtk} files by executing OpenFOAMs \texttt{foamToVTK} command, and a linear interpolator is generated and used for the initialization studies. 
%
%

\begin{table}[!ht]
\centering
\begin{tabular}{ l | l }
 \hline 
 software & OpenFOAM v10 \\
 \hline\hline
 Solver setting \\
 \hline
 solver, $p$  & PCG \\ 
 preconditioner, $p$ & DIC\\ 
 solver, $U$, $\tilde{\nu}$ & PBiCG \\ 
 preconditioner, $U$, $\tilde{\nu}$ & DILU\\ 
 tolerance &  $1e^{-10}$\\ 
 residual &   $1e^{-7}$\\ 
 relaxation factor & 0.3 \\ 
 \hline
 Numerical scheme setting \\
 \hline
 time scheme & steadyState\\ 
 gradient scheme & Gauss linear \\ 
 divergence scheme & linear \\ 
 div(phi,U)   & bounded Gauss linearUpwind grad(U)\\ 
 div(phi,nuTilda) & bounded Gauss upwind\\ 
 div((nuEff*dev2(T(grad(U))))) & Gauss linear\\ 
 Laplacian scheme & Gauss linear corrected\\ 
 interpolation scheme & linear \\ 
 surface normal gradient scheme & corrected\\ 
 wall distance scheme & meshWave \\ 
 \hline
\end{tabular}
\caption{Solver settings used in NACA0012 demonstration.}\label{tab:openfoam-setting}
\end{table}

\subsubsection{CNN wake extension model training}\label{sec:results-wake-extender}
\begin{figure}[htb!]
\centering
\includegraphics[width=4in]{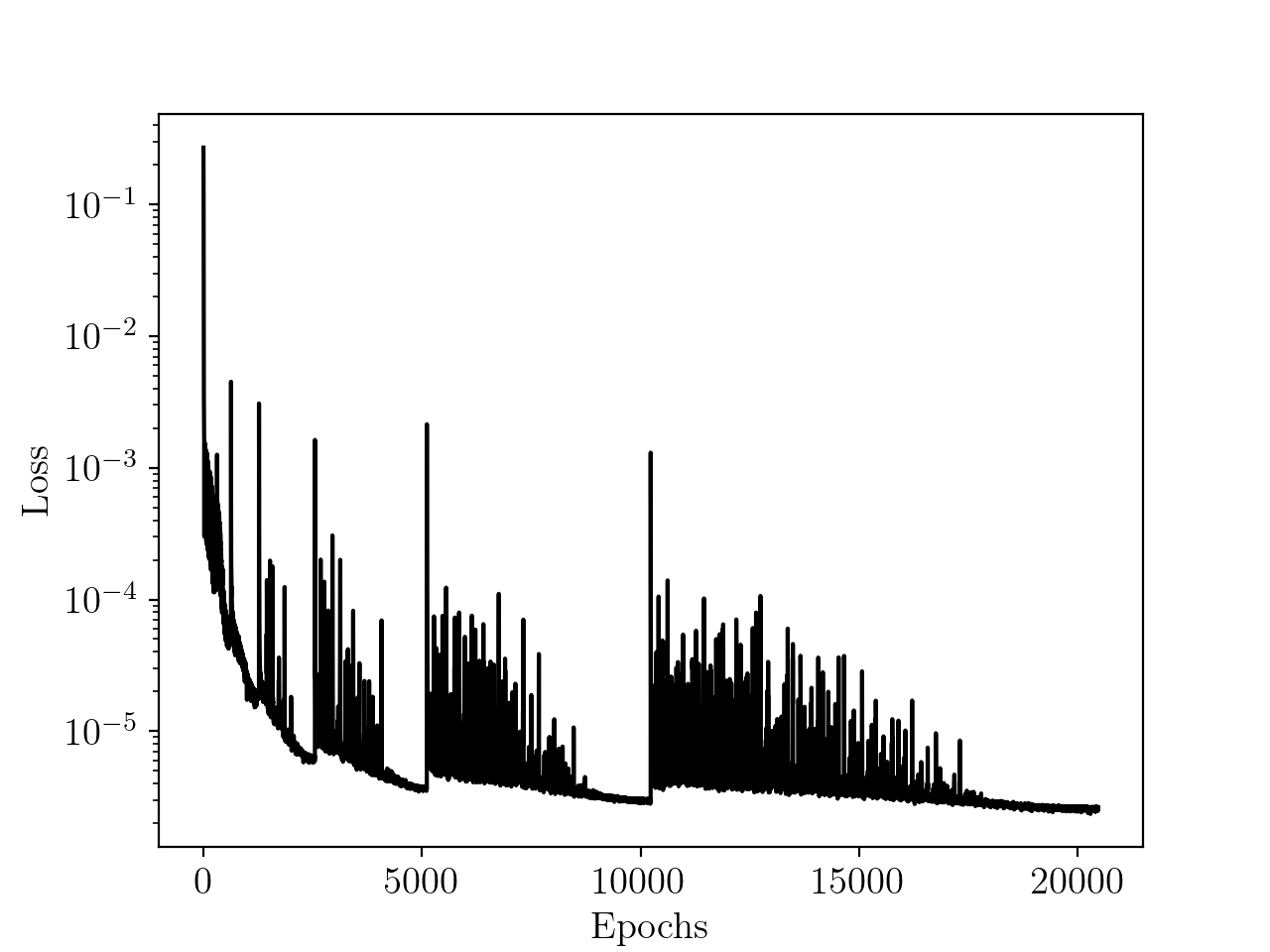}
\caption{Loss function evolution for the wake extension model training.}\label{fig:wake-loss}
\end{figure}

%
To prepare the CNN wake extension model training data, flow fields along $x-$slices at $x{=}x_{i}{=}2.0{+}i\Delta x$ for $i{=}0,..,98$, where $\Delta x {=} 1.0$, are evaluated at $n{=}128$ equally spaced points along the $y-$direction in the range $\left[y_{min},y_{max}\right]=\left[-1.0,1.0\right]$.
The CNN model is  constructed as illustrated in Fig. \ref{fig:cnn-architecture} with kernel size 2 for the convolution and max pooling layers. The activation layers use SiLU activation function. AdamW optimizer with cosine annealing learning rate scheduling \cite{Loshchilov2017} is used to train the model. In cosine annealing, the learning rate of the optimizer is increased and then slowly decreased in repeated intervals to warm start the training process multiple times.
Batch optimization with batch size 8 is used. Parameters used in the data generation, CNN architecture properties, and model training optimizer setting are summarized in Tabs. \ref{tab:cnn-param-data}, \ref{tab:cnn-param-model}, and \ref{tab:cnn-param-training}, respectively. The evolution of the loss function over epochs during training is shown in Fig. \ref{fig:wake-loss}.
The loss function jumps up to higher values a few times throughout the entire training process due to the sudden increase in the learning rate as set by the cosine annealer. The maximum number of epochs is set to the end of the annealing cycle. 

\begin{table}[!ht]
\centering
\begin{tabular}{ l | l }
 \hline 
 Training data parameter & Value\\
 \hline 
 starting slice location $x_{0}$& 2.0 \\ 
 slice distance $\Delta x$ & 1.0\\ 
 number of slices $m$ & 98 \\ 
 number of $y$ points $n$ & 128 \\ 
 \hline
\end{tabular}
\caption{CNN model training data parameters.}\label{tab:cnn-param-data}
\end{table}

\begin{table}[!ht]
\centering
\begin{tabular}{ l | l }
 \hline 
 CNN model architecture property & Name/value\\
 \hline 
 activation function & SiLU \\
 conv/max pooling layer repetition &3\\
 convolutional layer kernel size& 2\\ 
 max pooling kernel size& 2\\
 stride & 1\\
 padding & 0\\
 dilation & 1\\
 \hline
\end{tabular}
\caption{CNN model architecture properties.}\label{tab:cnn-param-model}
\end{table}

\begin{table}[!ht]
\centering
\begin{tabular}{ l | l }
 \hline 
 Optimizer setting & Name/value\\
 \hline 
 optimizer & AdamW \\\
 $\beta_{1}$ & 0.9 (default)\\
 $\beta_{2}$ & 0.999 (default)\\
 weight decay & 0.01 (default)\\
 batch size & 8\\ 
 learning rate scheduling & cosine annealing \\
 $T_{0}$ &10\\
 $T_{mult}$ & 2\\
 $\eta_{min}$ & $1e^{-6}$\\
 initial learning rate & $1e^{-3}$\\ 
 \hline
\end{tabular}
\caption{CNN model training optimizer setting.}\label{tab:cnn-param-training}
\end{table}

\begin{figure}[htb!]
    \centering
    \begin{subfigure}{2.6in}
        \includegraphics[width=\textwidth]{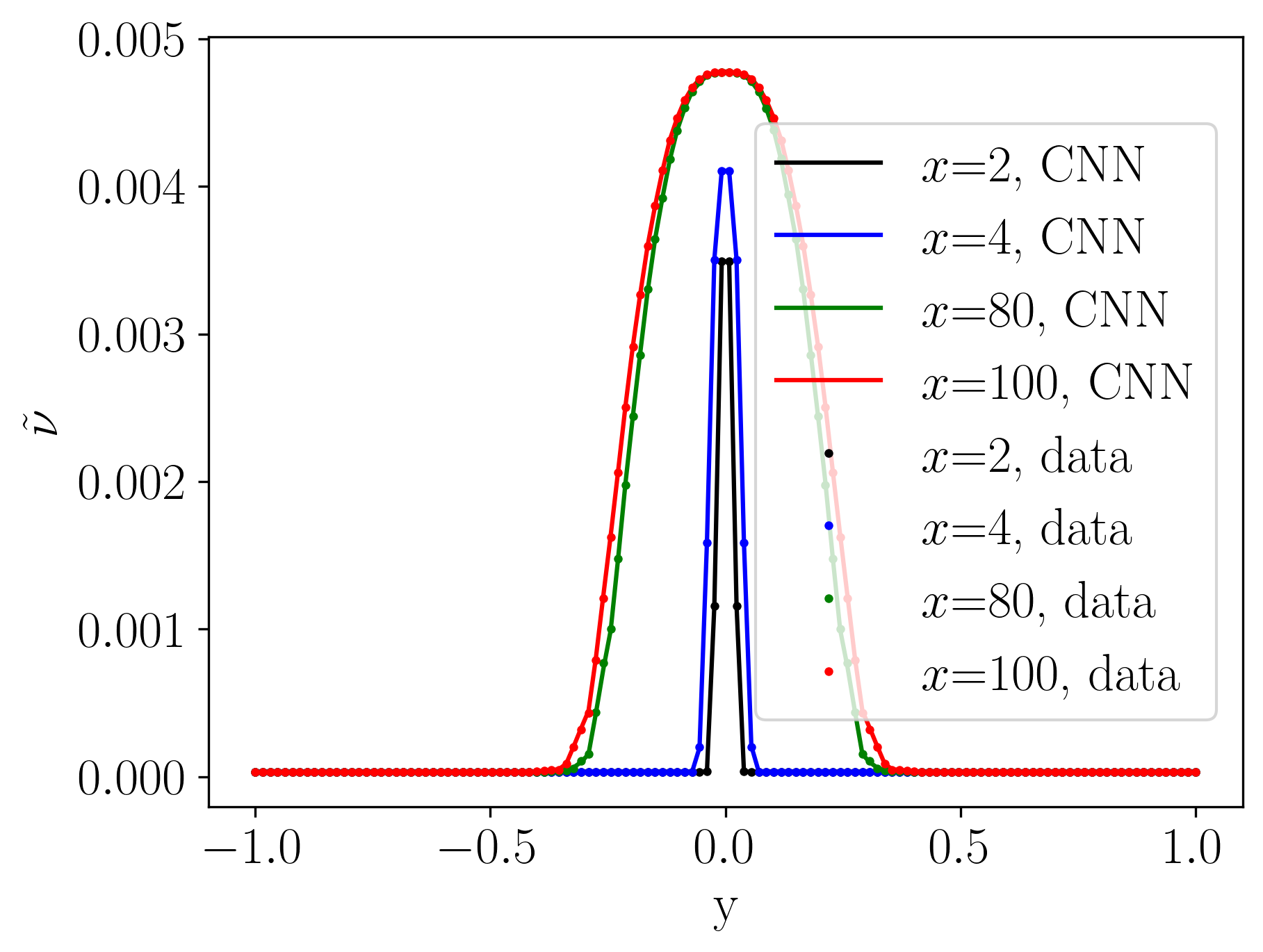}
        \caption{$\tilde{\nu}$}
    \end{subfigure}
    \begin{subfigure}{2.6in}
        \includegraphics[width=\textwidth]{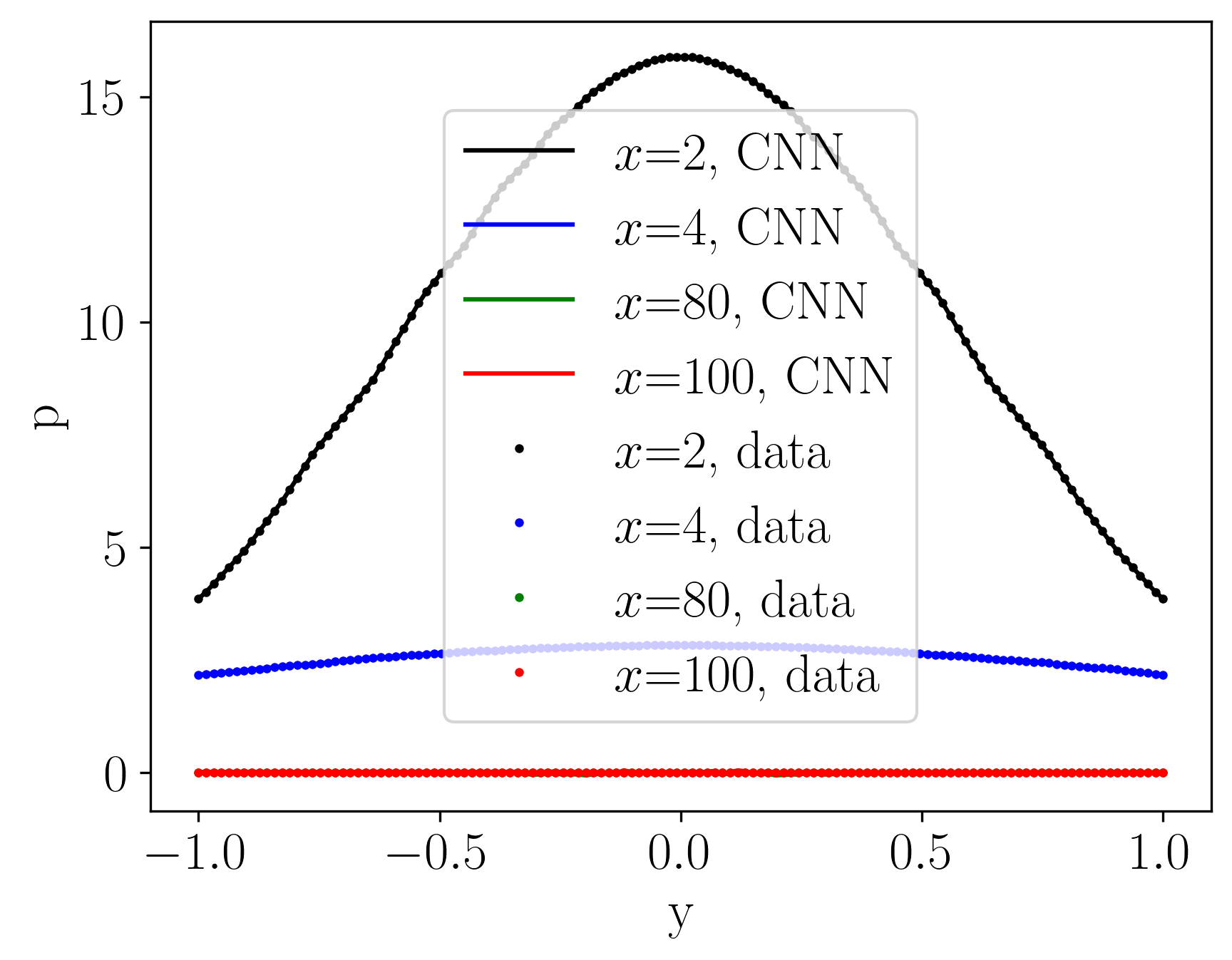}
        \caption{$p$}
    \end{subfigure}
    \\
        \begin{subfigure}{2.6in}
        \includegraphics[width=\textwidth]{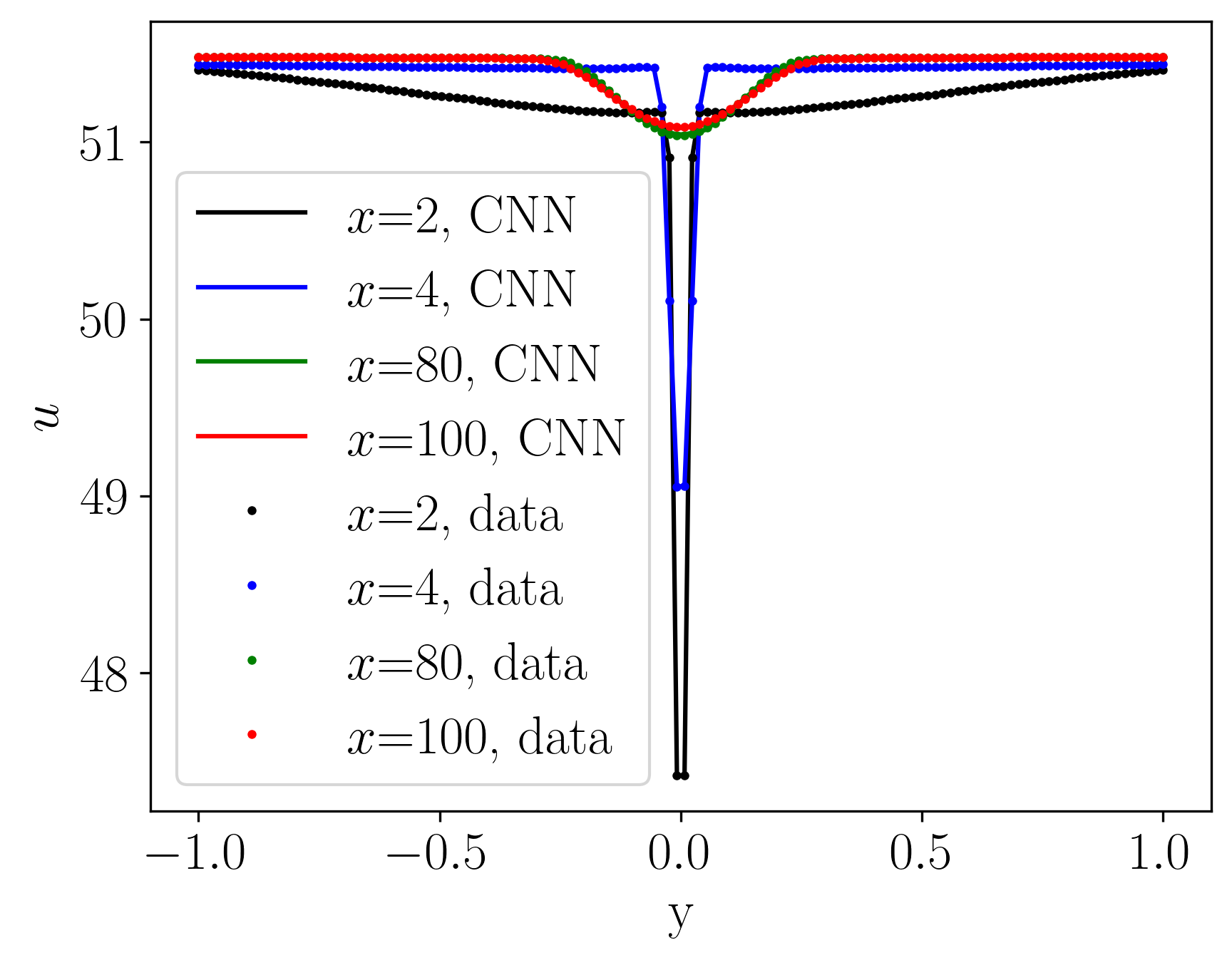}
        \caption{$u$}
    \end{subfigure} 
    \begin{subfigure}{2.6in}
        \includegraphics[width=\textwidth]{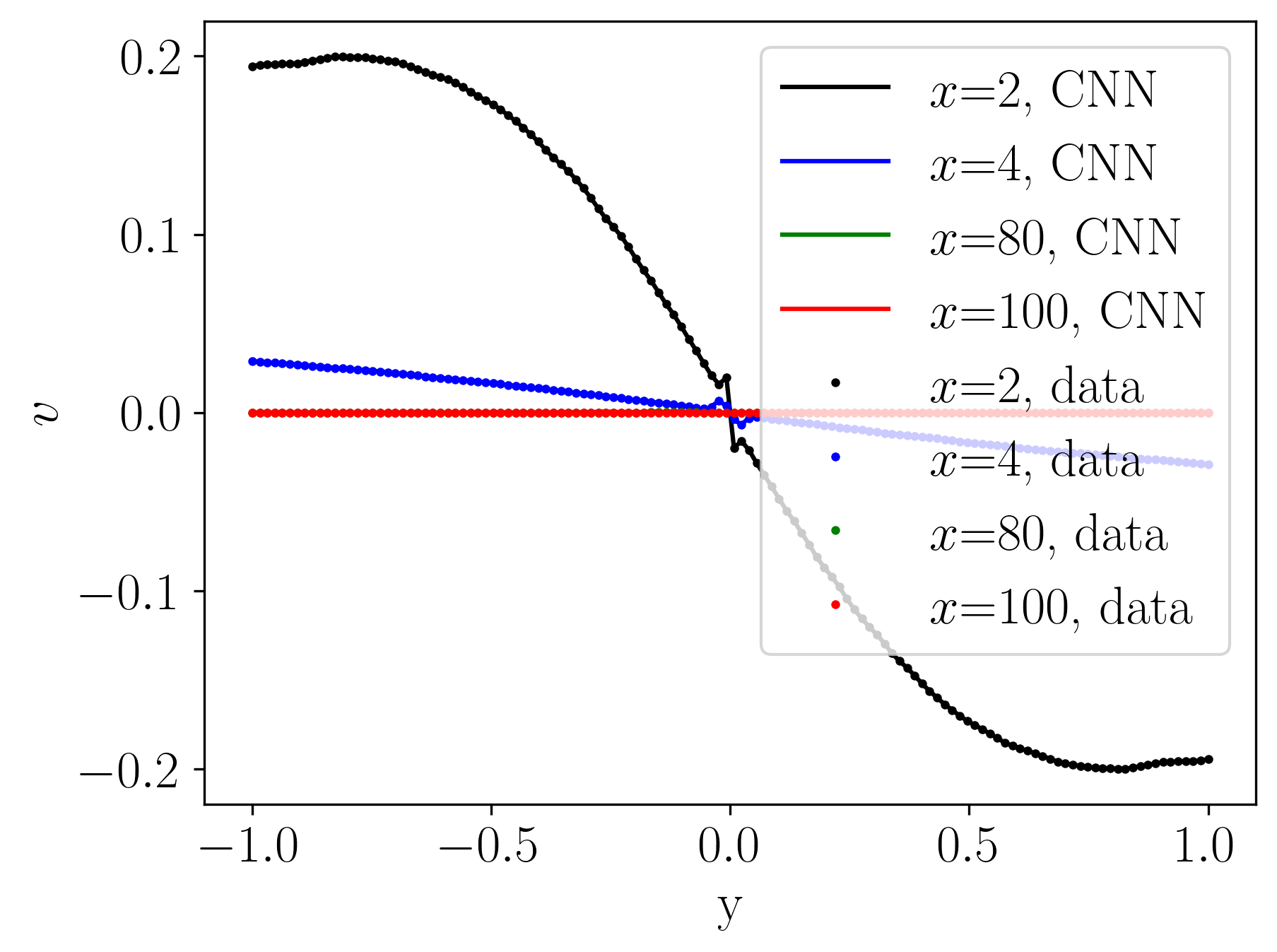}
        \caption{$v$}
    \end{subfigure}
    \caption{Flow field $y-$profiles of the wake, compared between prediction by the CNN model (solid) and data (dots).}\label{fig:wake-cross-section}
\end{figure}

The trained CNN model can be used to reproduce the wake in the entire wake region, down to the right end of the computational domain, following Algorithm \ref{alg:cnn-wake}. The location of the near-body/wake transition interface, $x_{interface}$, can be set retroactively at any $x \geq x_{0}$, however, the $y-$range, $\left[y_{min},y_{max}\right]$, number of evaluation points, $n$, and slice spacing, $\Delta x$, cannot be changed after the CNN model is trained. For the CFD initialization studies in Sec. \ref{sec:results-init}, the interface location $x_{interface}$ is varied during the parametric study. 
Using $x=x_{0}=2.0$ as the starting slice, predicted $y-$profiles of the flow features at several $x-$locations are plotted in solid curves alongside the data from the HF solution in dots in Fig. \ref{fig:wake-cross-section}. 
While all the five flow fields are modeled, Spalart-Allmaras viscosity $\tilde{\nu}$ and turbulent kinematic viscosity $\nu_{t}$ become identical away from the wall, and only the wake profiles for $\tilde{\nu}$ are shown here for brevity. 
These plots highlight the various ways in which the physics of wake development presents itself in different field quantities. For instance, the Spalart-Allmaras viscosity, $\tilde{\nu}$ (see Fig. \ref{fig:wake-cross-section}(a)), and the component of the velocity along the flow direction, $u$ (see Fig. \ref{fig:wake-cross-section}(c)), have salient features that persist far downstream in the wake, while the other field quantities decay quickly to freestream values. The trained CNN models these features well, as indicated in Fig. \ref{fig:wake-cross-section} throughout the wake region, all the way to the end of the analysis domain at $x=100$.

\begin{figure}[h!]
    \centering
    \begin{subfigure}{0.24\textwidth}
        \includegraphics[width=\textwidth]{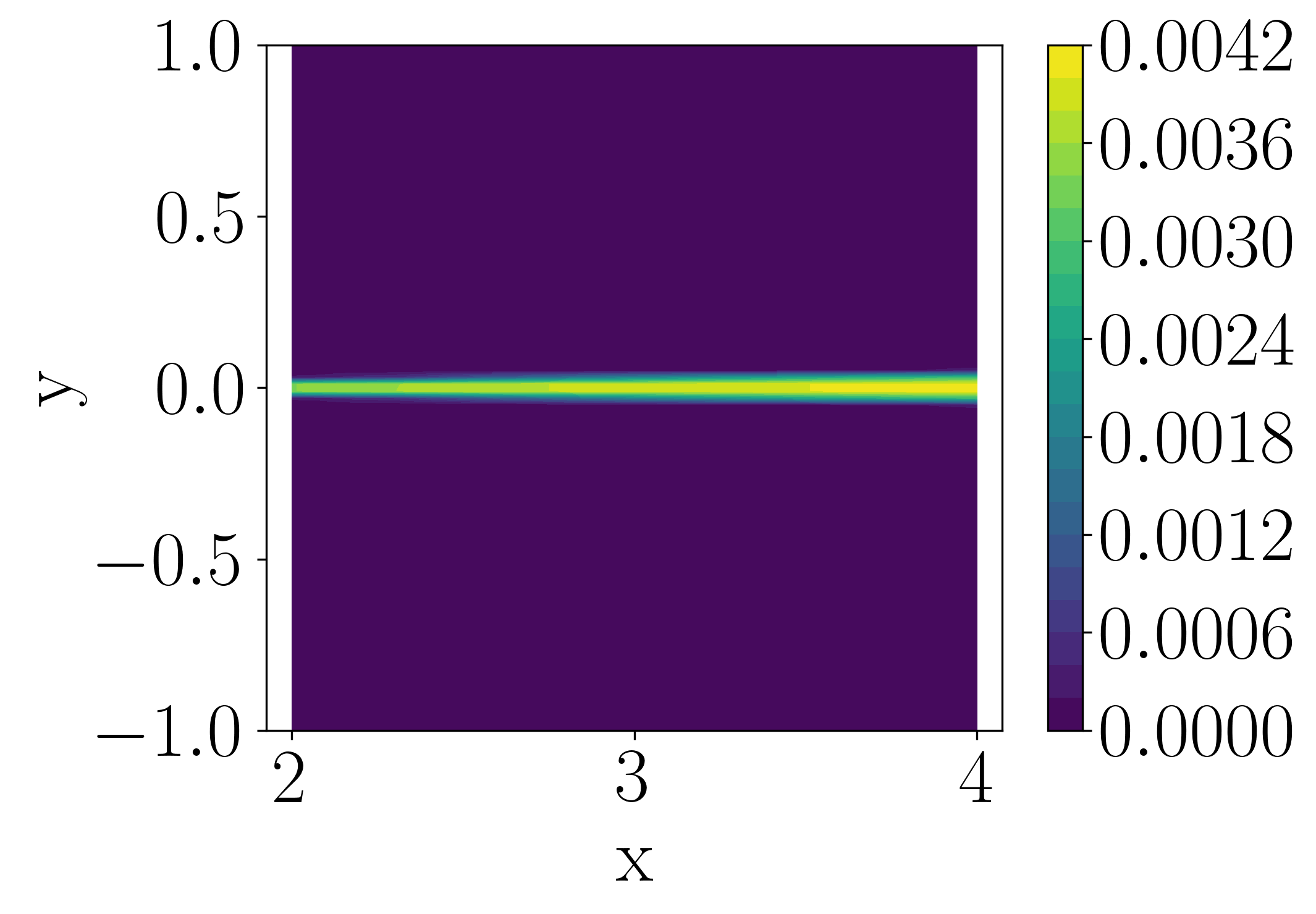}
        \caption{HF $\tilde{\nu}$}
    \end{subfigure}
    \begin{subfigure}{0.24\textwidth}
        \includegraphics[width=\textwidth]{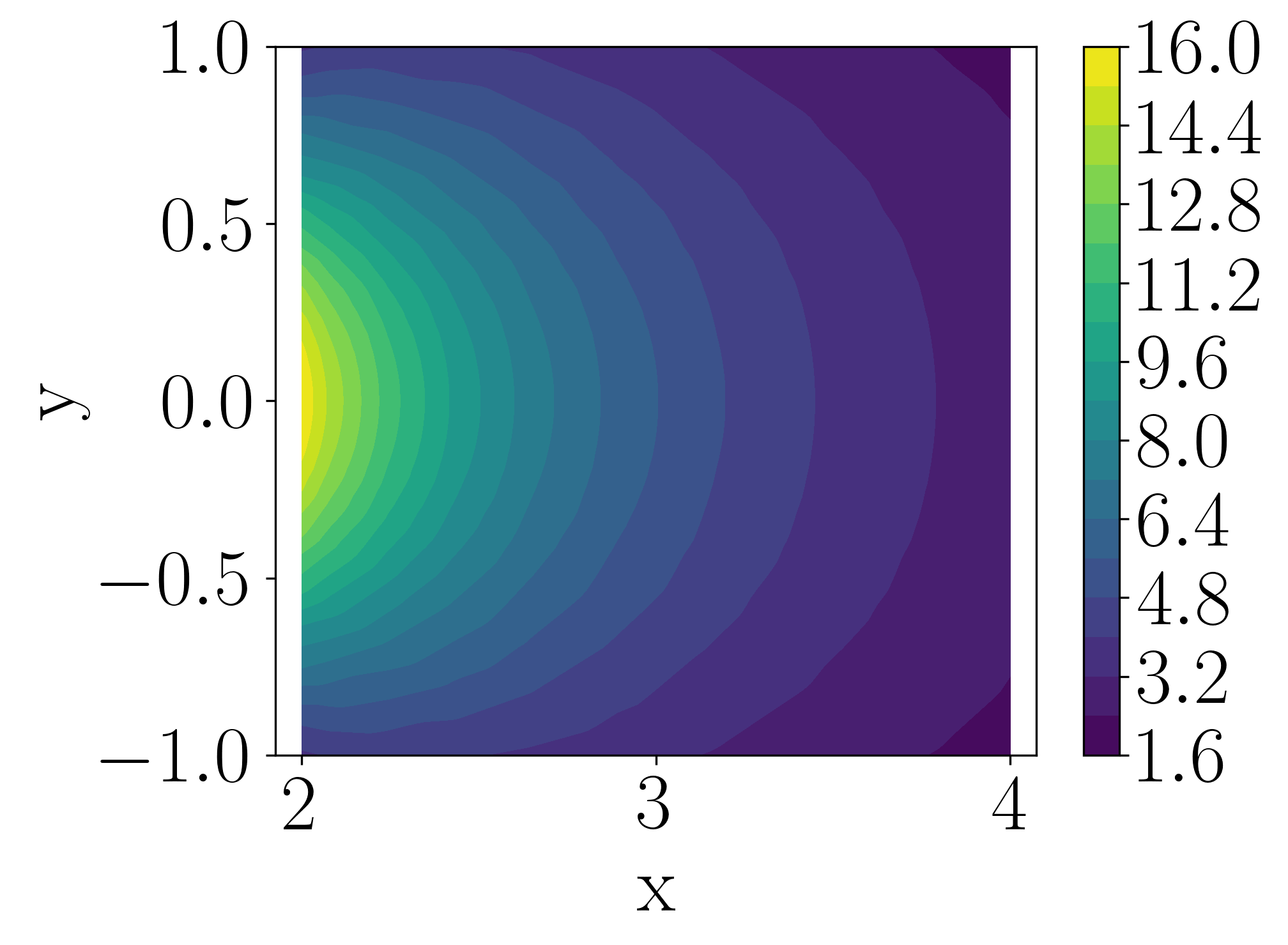}
    \caption{HF $p$}
    \end{subfigure}
    \begin{subfigure}{0.24\textwidth}
        \includegraphics[width=\textwidth]{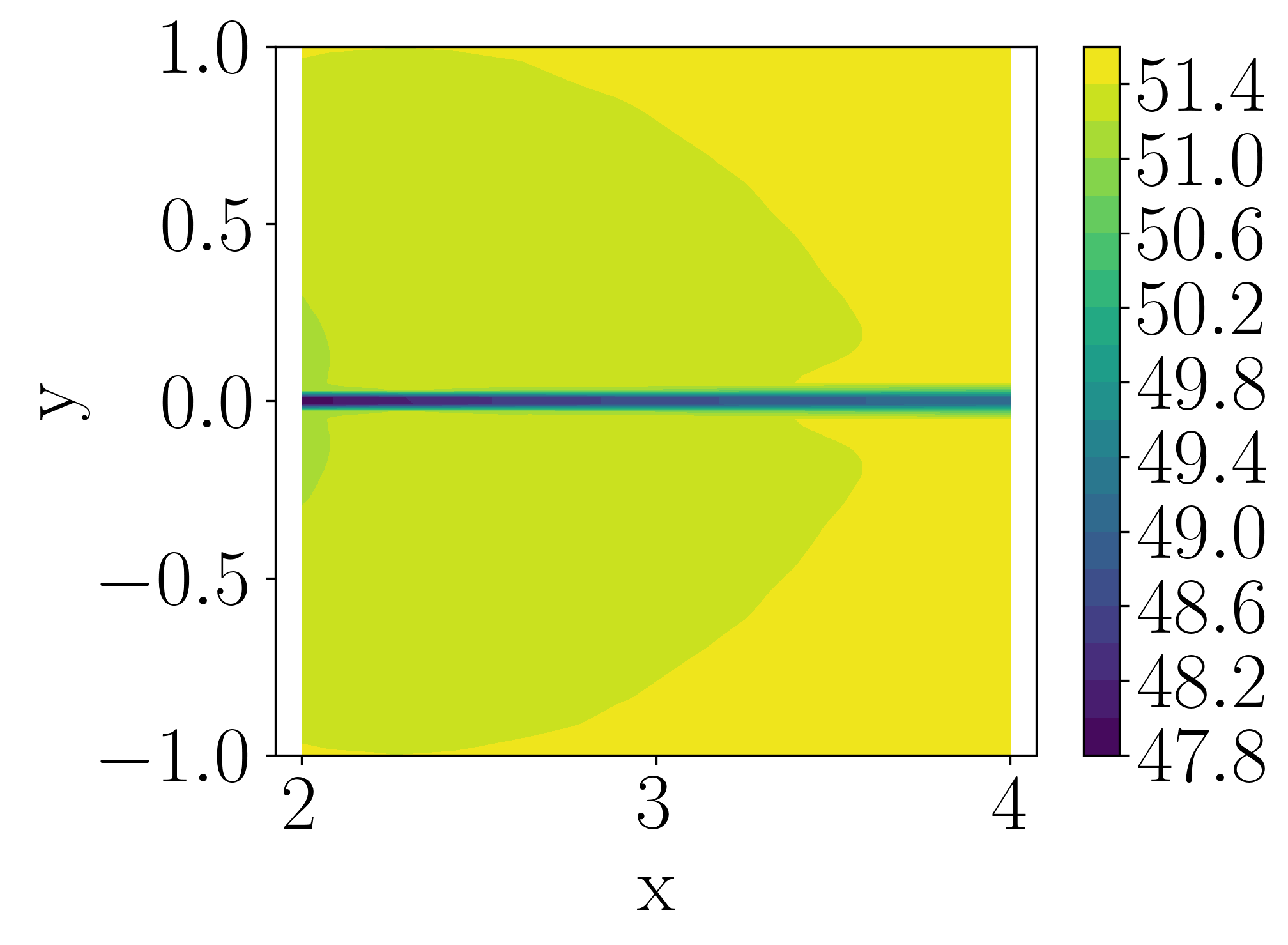}
        \caption{HF $u$}
    \end{subfigure}
    \begin{subfigure}{0.24\textwidth}
        \includegraphics[width=\textwidth]{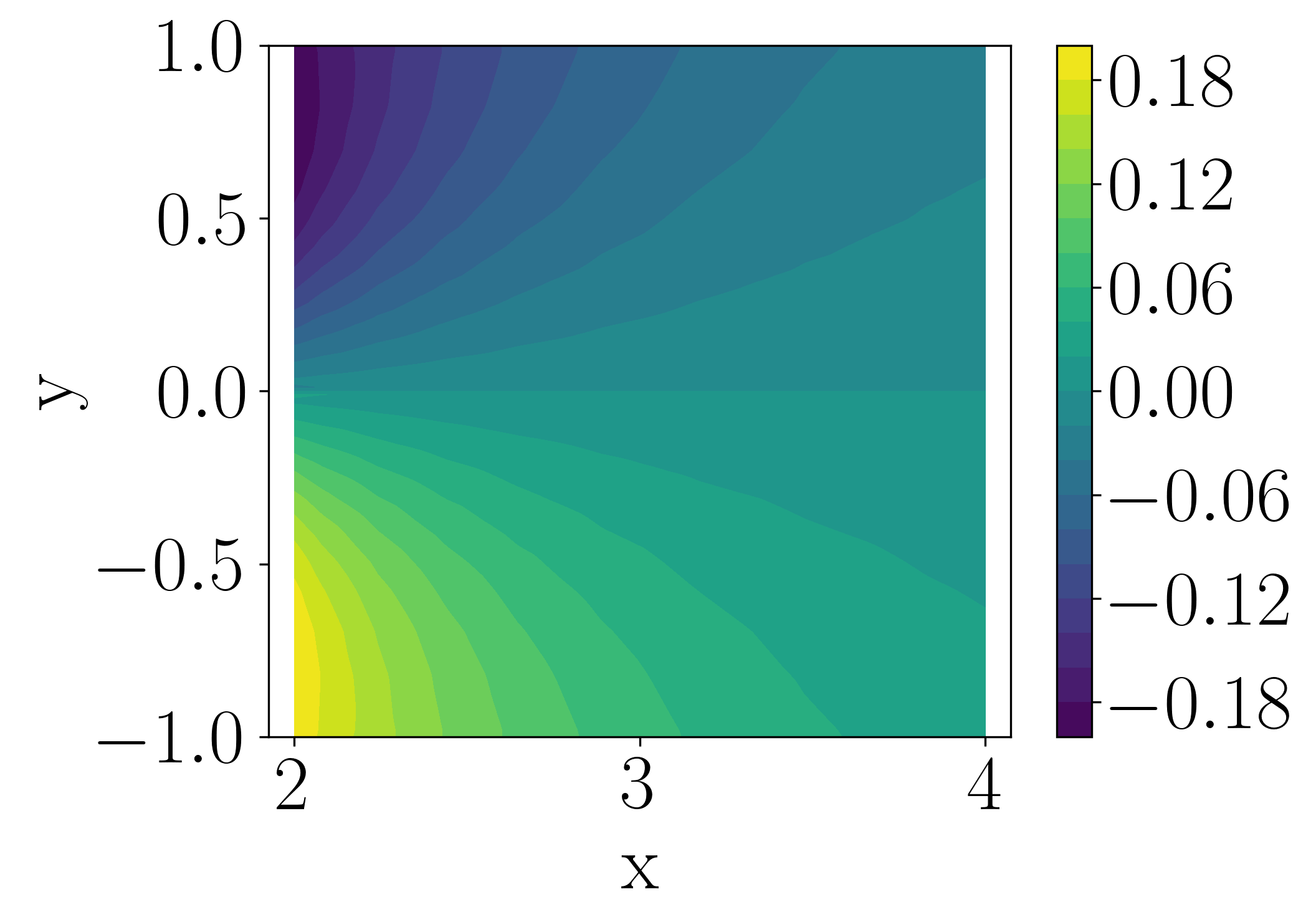}
        \caption{HF $v$}
    \end{subfigure}
    \\
    \begin{subfigure}{0.24\textwidth}
        \includegraphics[width=\textwidth]{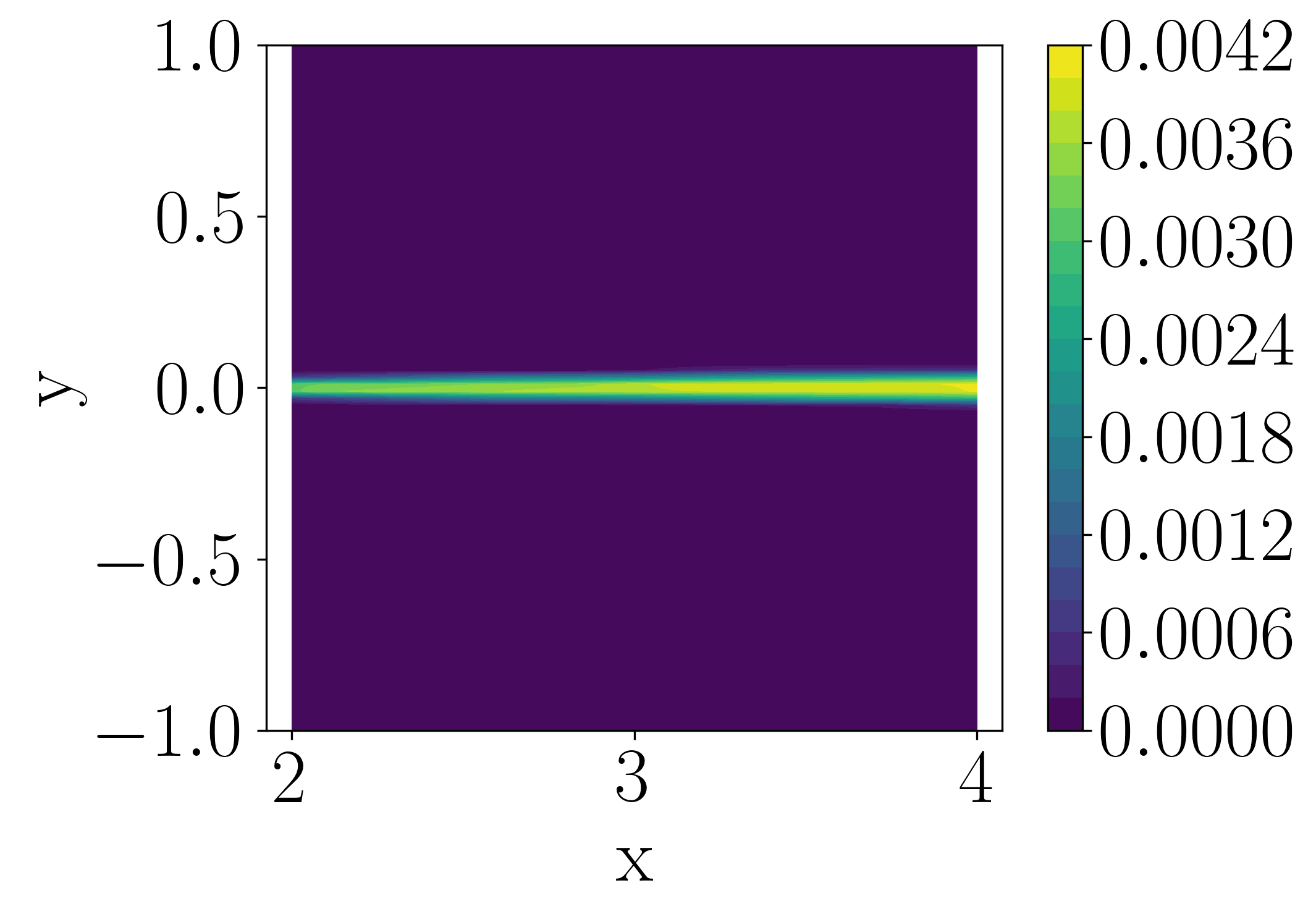}
        \caption{CNN $\tilde{\nu}$}
    \end{subfigure}
    \begin{subfigure}{0.24\textwidth}
        \includegraphics[width=\textwidth]{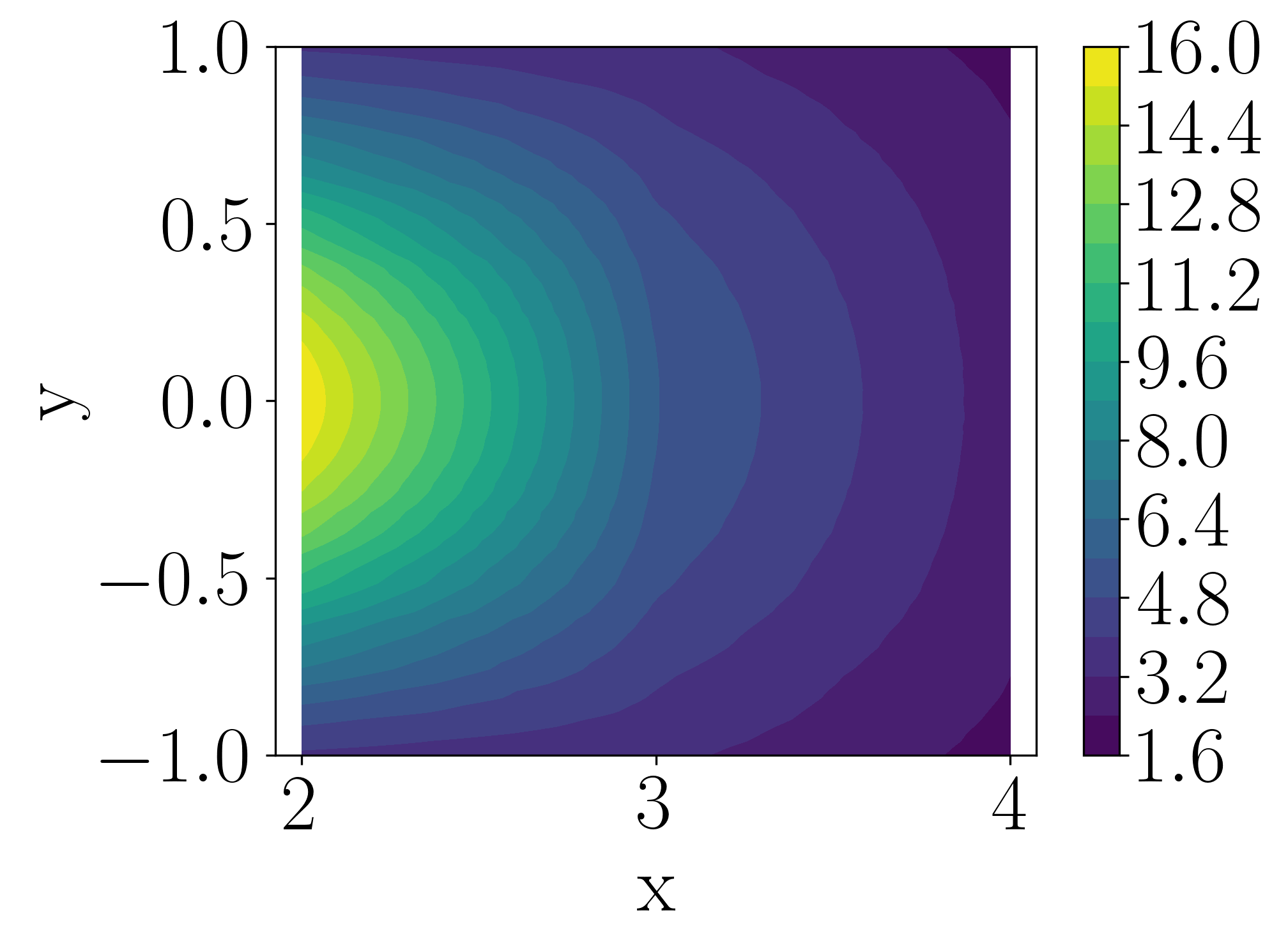}
    \caption{CNN $p$}
    \end{subfigure}
    \begin{subfigure}{0.24\textwidth}
        \includegraphics[width=\textwidth]{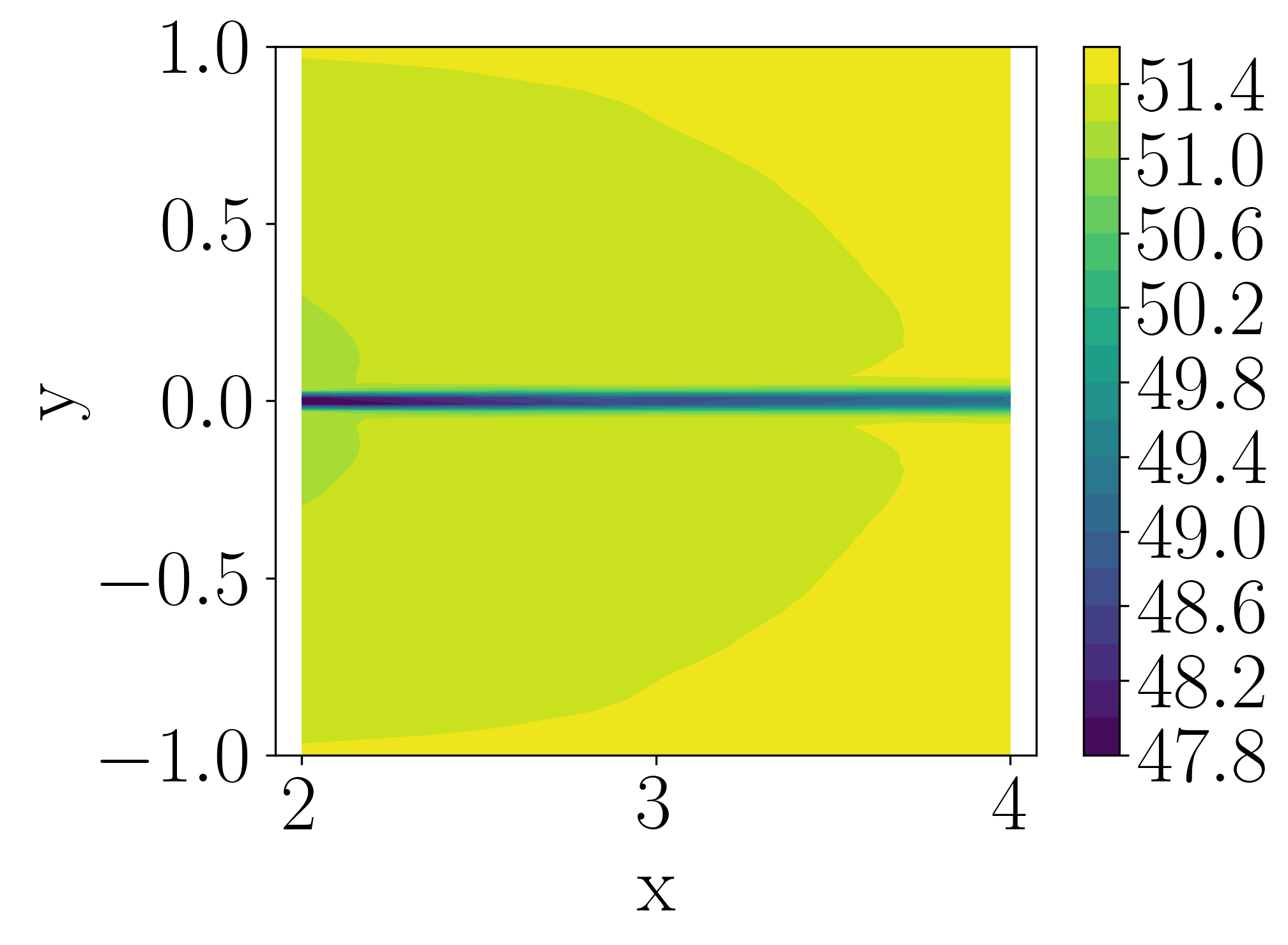}
        \caption{CNN $u$}
    \end{subfigure}
    \begin{subfigure}{0.24\textwidth}
        \includegraphics[width=\textwidth]{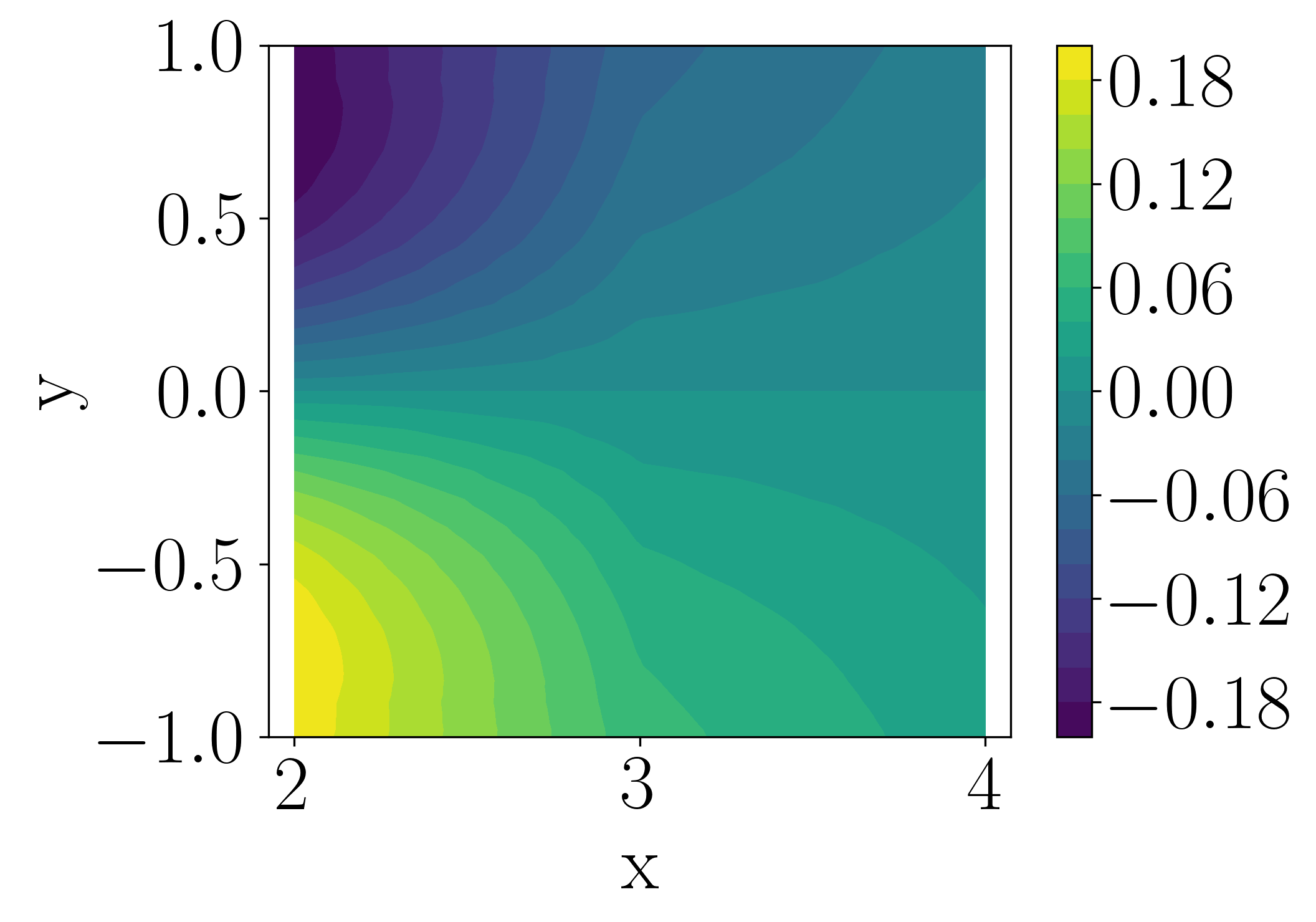}
        \caption{CNN $v$ }
    \end{subfigure}
    \\
    \begin{subfigure}{0.24\textwidth}
        \includegraphics[width=\textwidth]{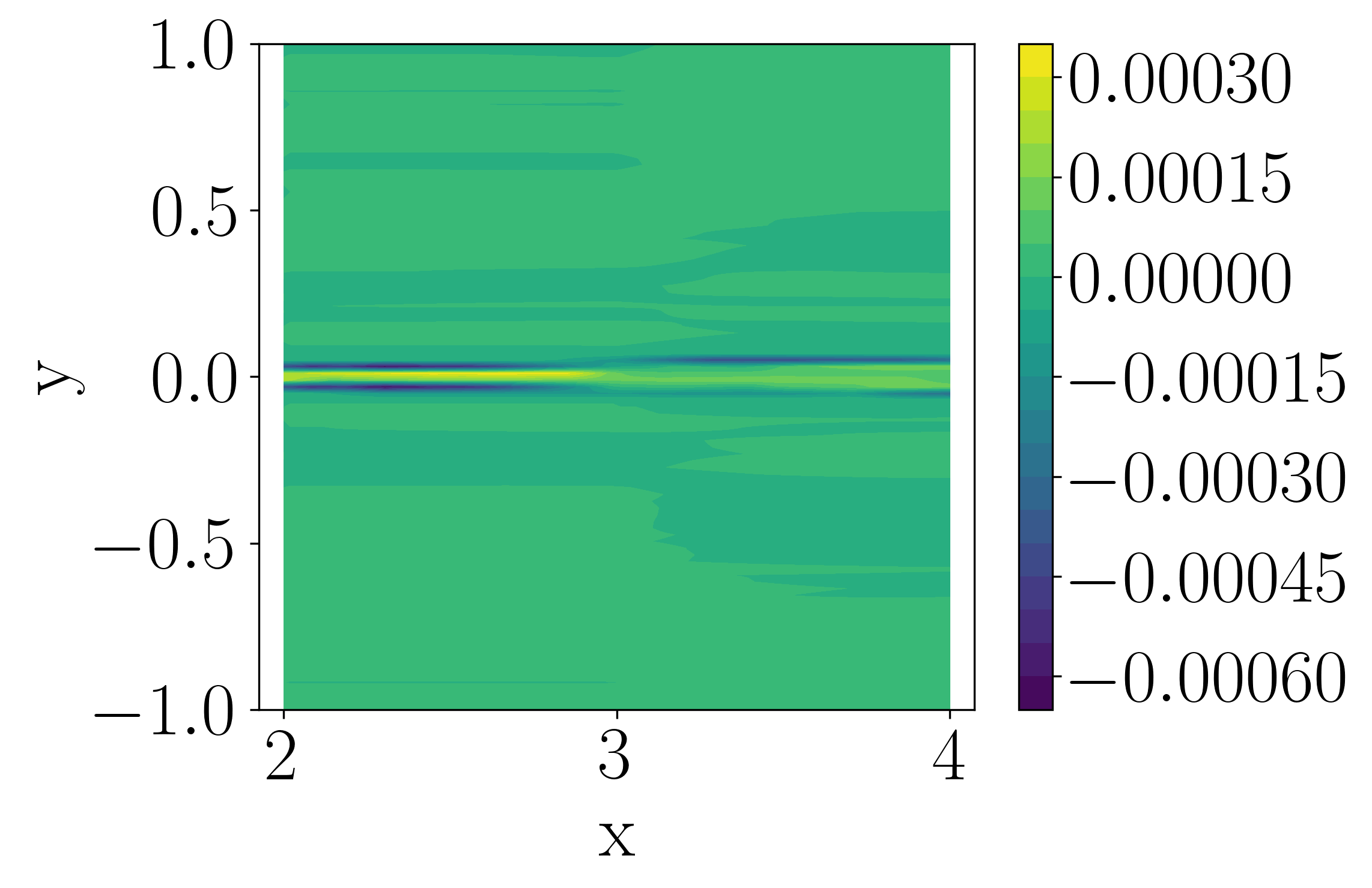}
        \caption{Error in $\tilde{\nu}$}
    \end{subfigure}
    \begin{subfigure}{0.24\textwidth}
        \includegraphics[width=\textwidth]{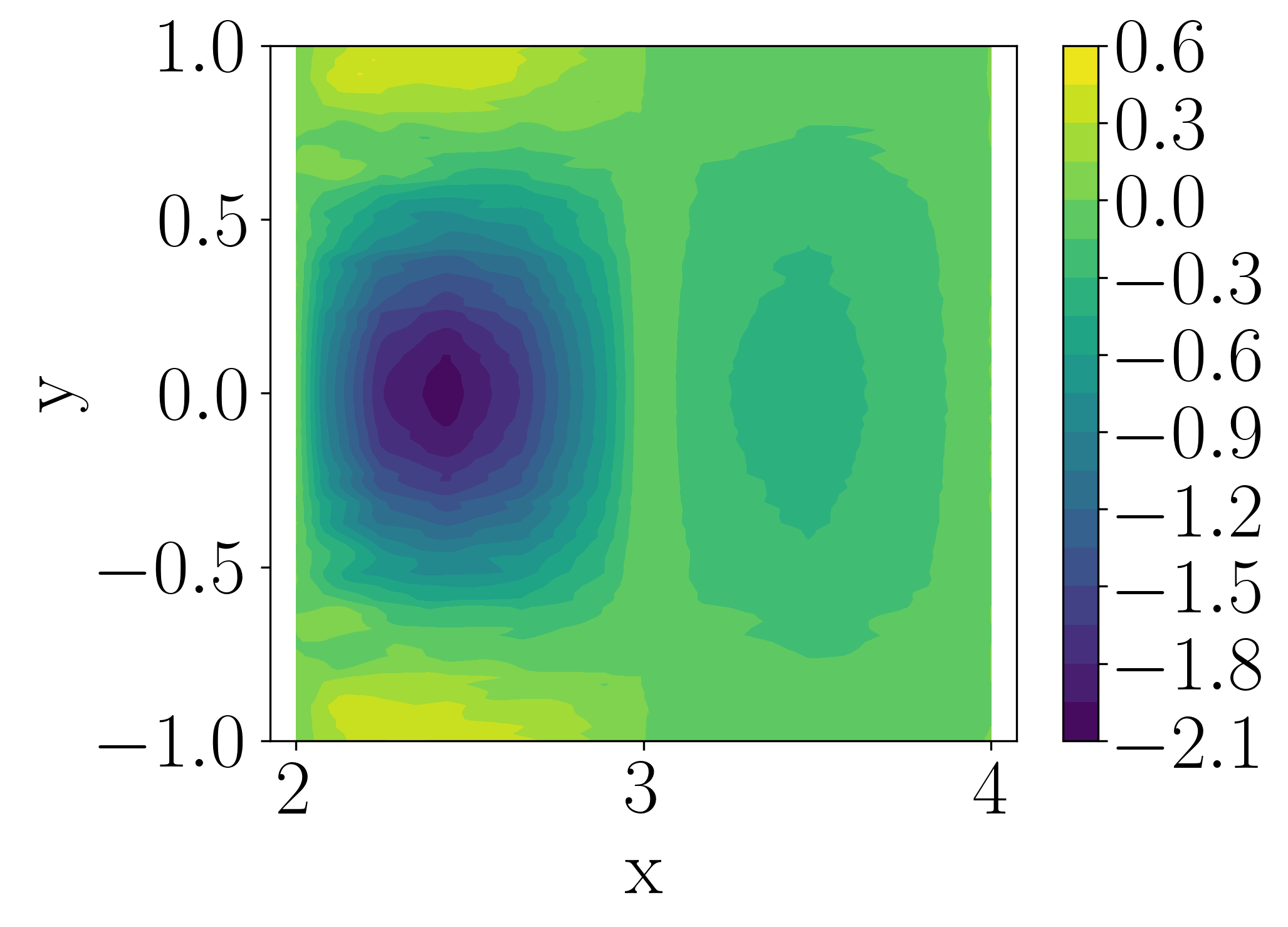}
    \caption{Error in $p$}
    \end{subfigure}
    \begin{subfigure}{0.24\textwidth}
        \includegraphics[width=\textwidth]{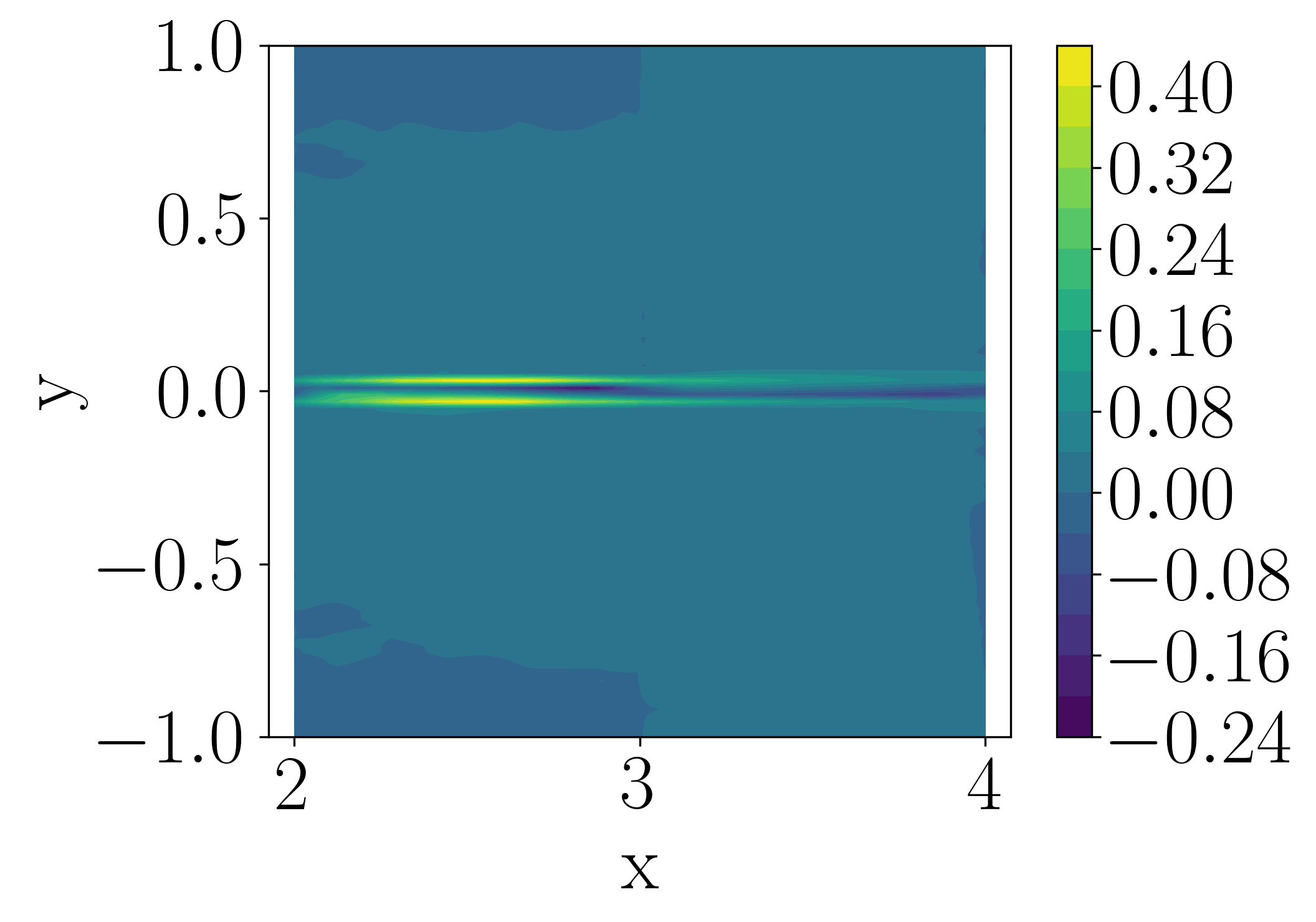}
        \caption{Error in $u$}
    \end{subfigure}
    \begin{subfigure}{0.24\textwidth}
        \includegraphics[width=\textwidth]{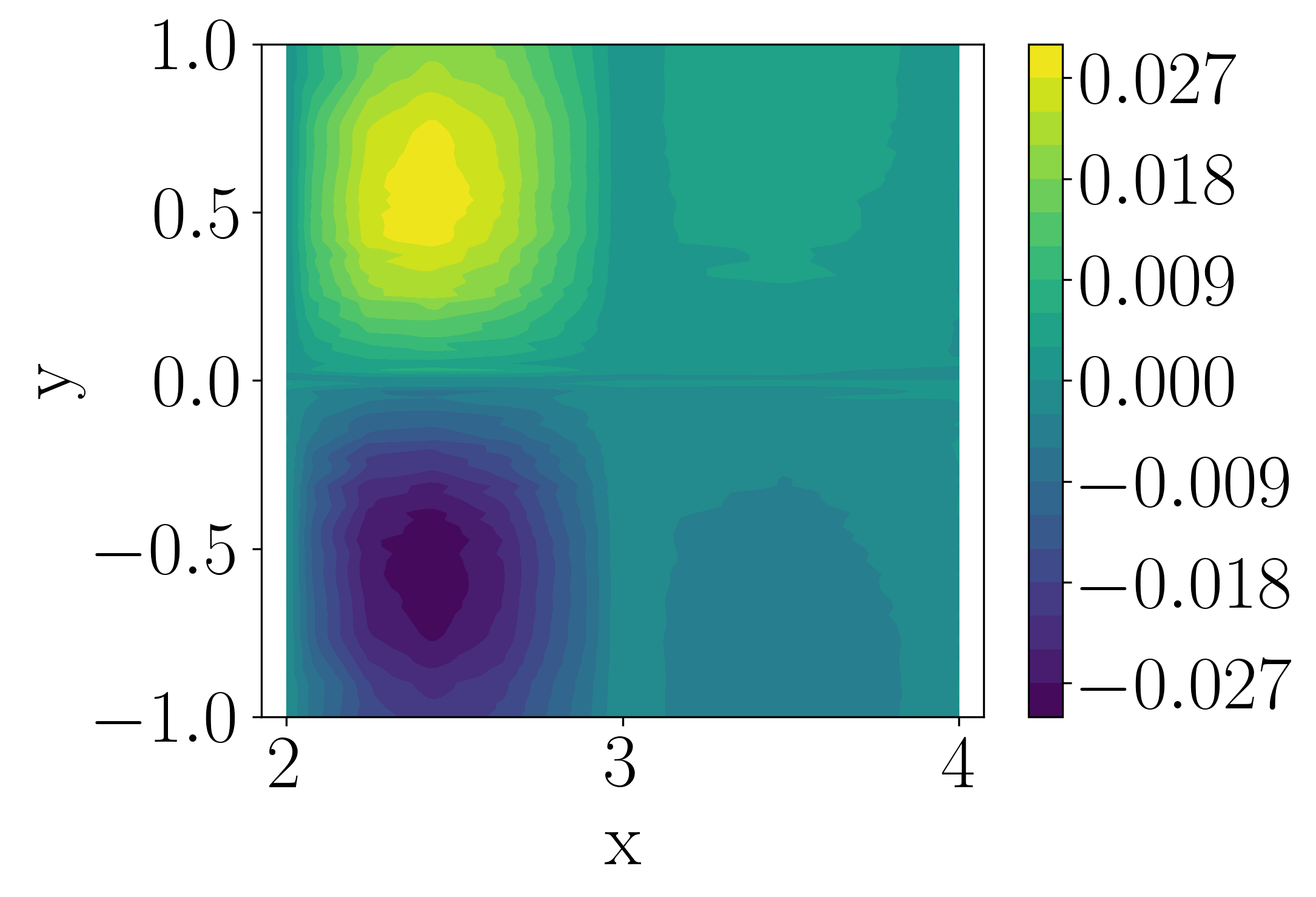}
        \caption{Error in $v$}
    \end{subfigure}
    \caption{Comparison between HF and CNN-predicted wake at $x\in\left[2.0,4.0\right]$.}\label{fig:cnn-result-2}
\end{figure}

\begin{figure}[h!]
    \centering
   \begin{subfigure}{0.24\textwidth}
        \includegraphics[width=\textwidth]{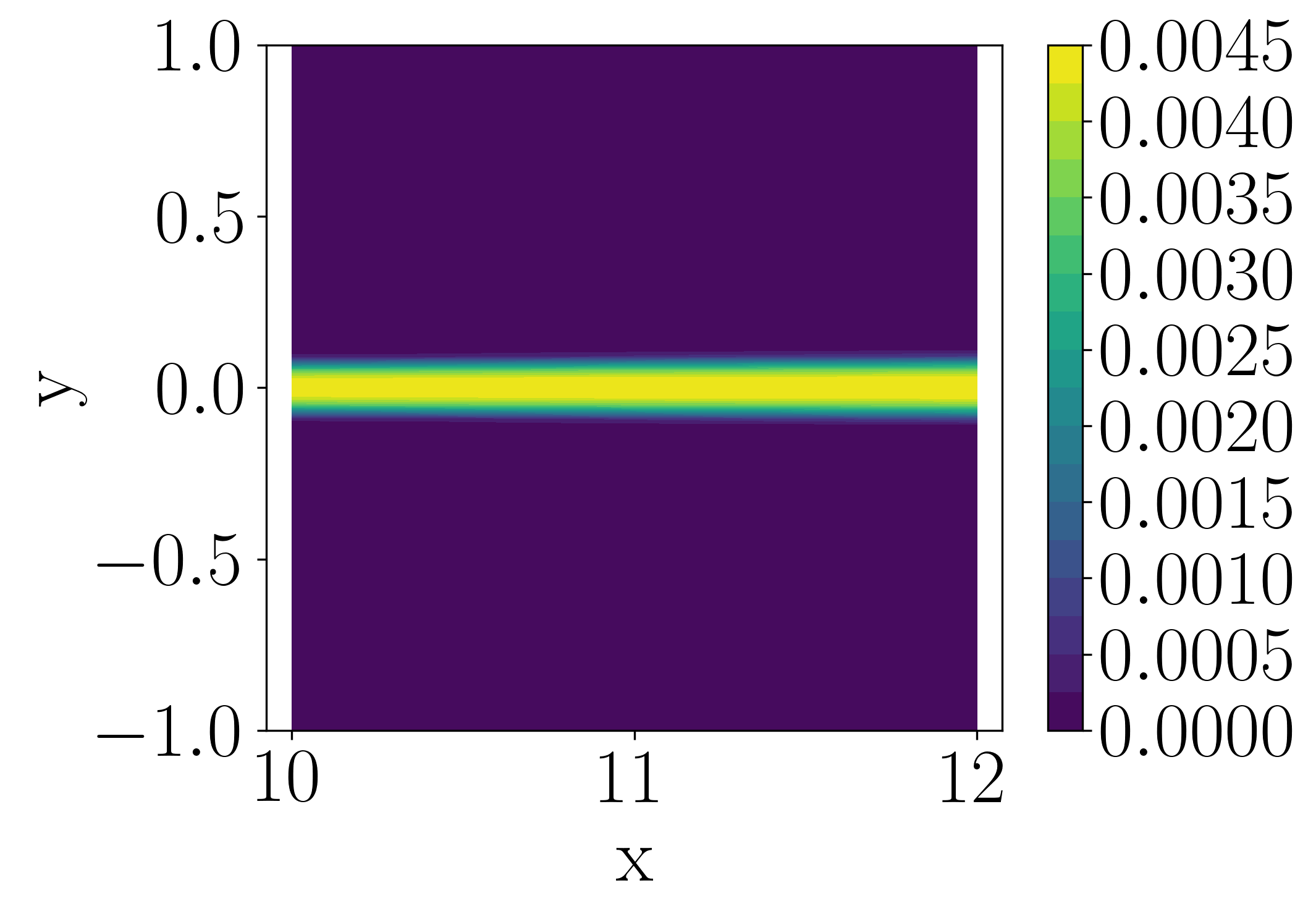}
        \caption{HF $\tilde{\nu}$}
    \end{subfigure}
    \begin{subfigure}{0.24\textwidth}
        \includegraphics[width=\textwidth]{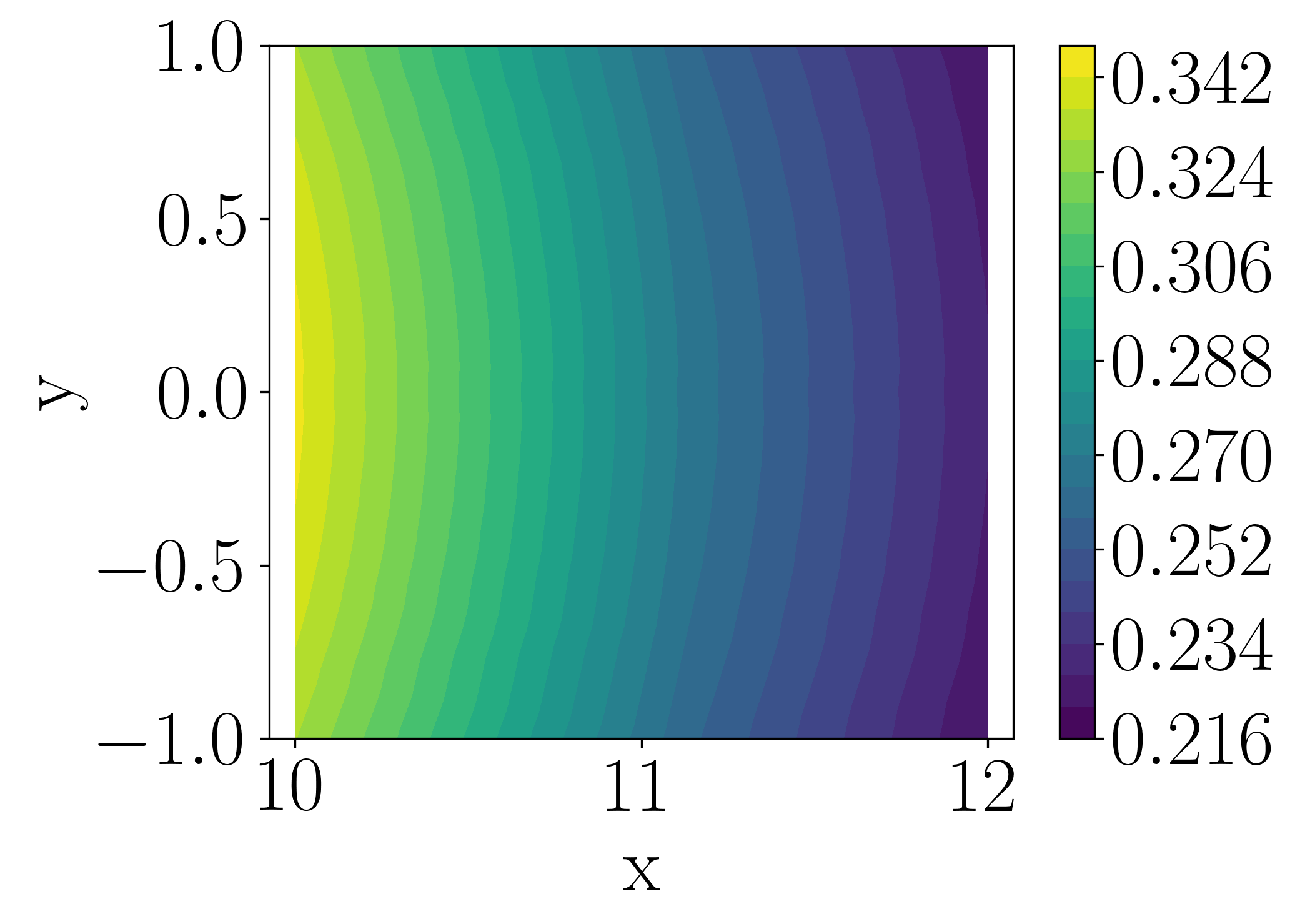}
    \caption{HF $p$}
    \end{subfigure}
    \begin{subfigure}{0.24\textwidth}
        \includegraphics[width=\textwidth]{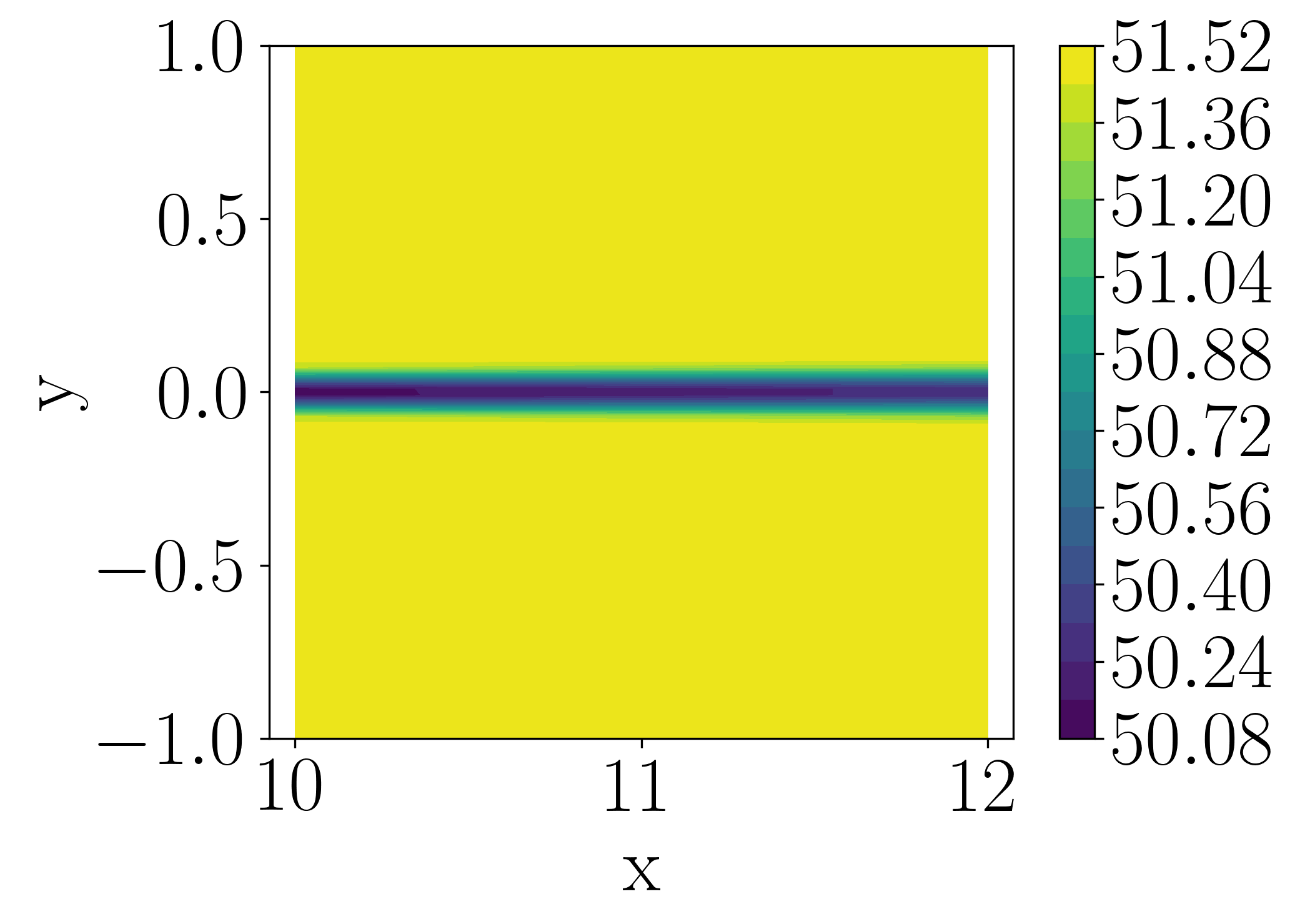}
        \caption{HF $u$}
    \end{subfigure}
    \begin{subfigure}{0.24\textwidth}
        \includegraphics[width=\textwidth]{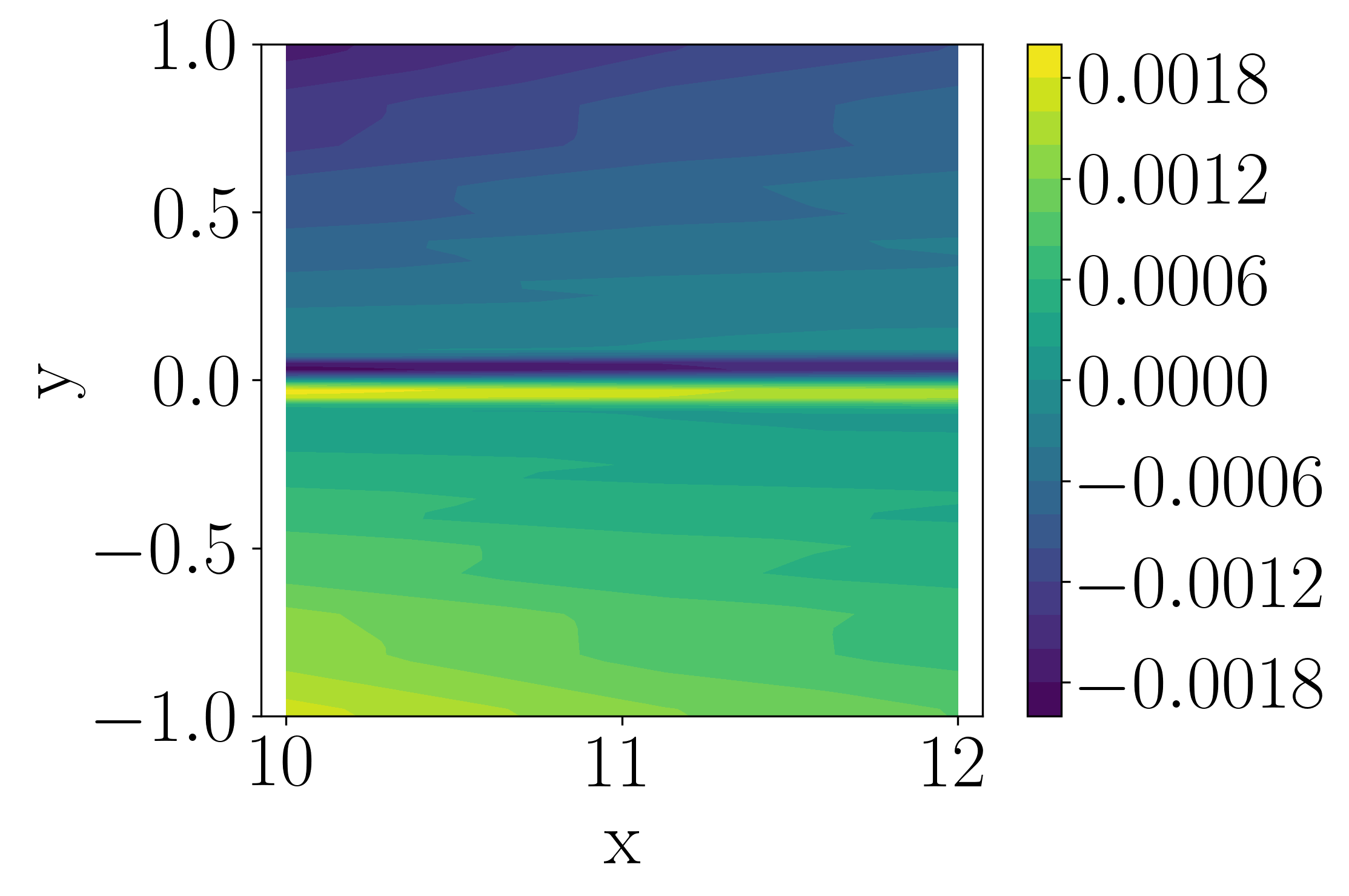}
        \caption{HF $v$}
    \end{subfigure}
    \\
   \begin{subfigure}{0.24\textwidth}
        \includegraphics[width=\textwidth]{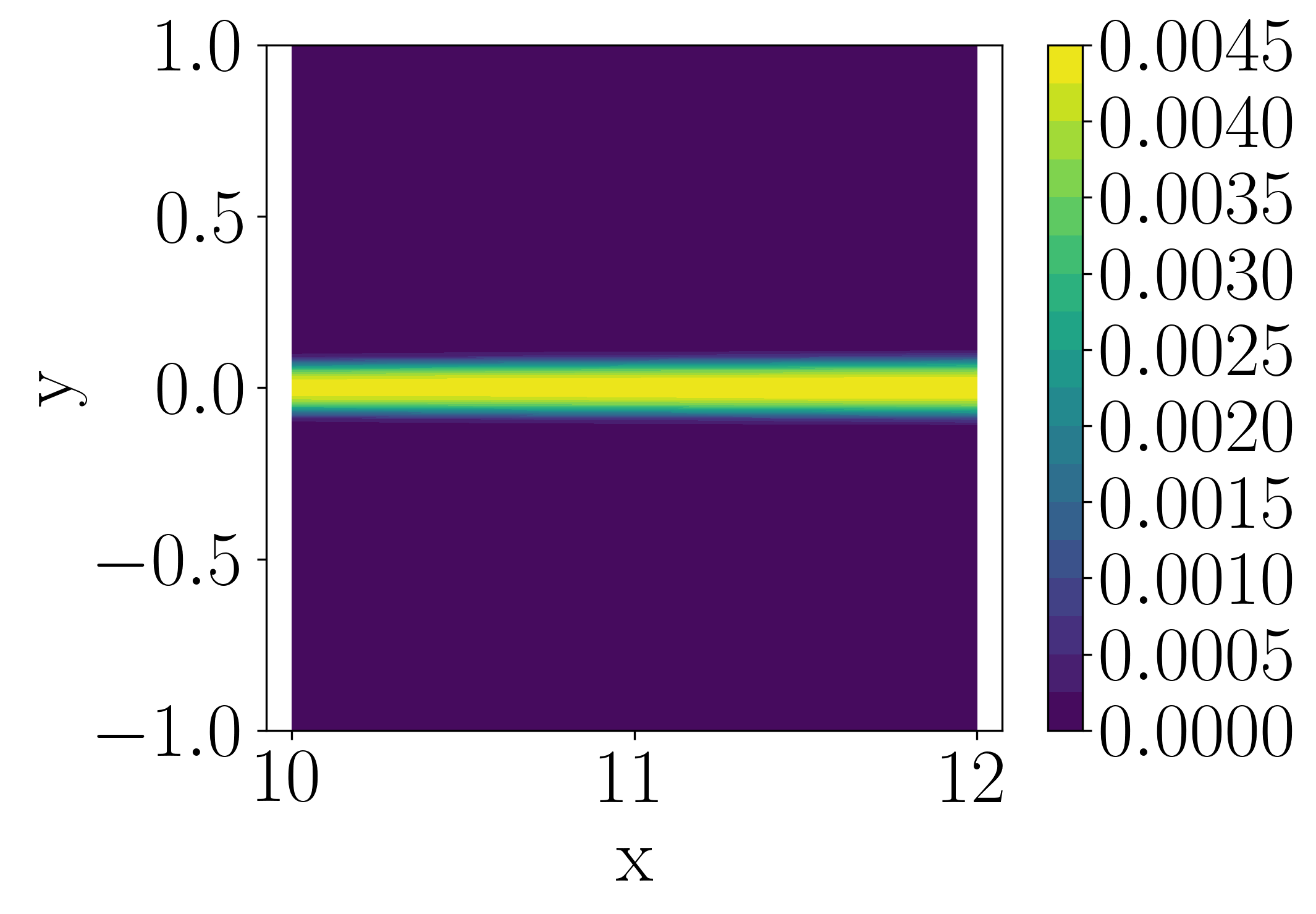}
        \caption{CNN $\tilde{\nu}$}
    \end{subfigure}
    \begin{subfigure}{0.24\textwidth}
        \includegraphics[width=\textwidth]{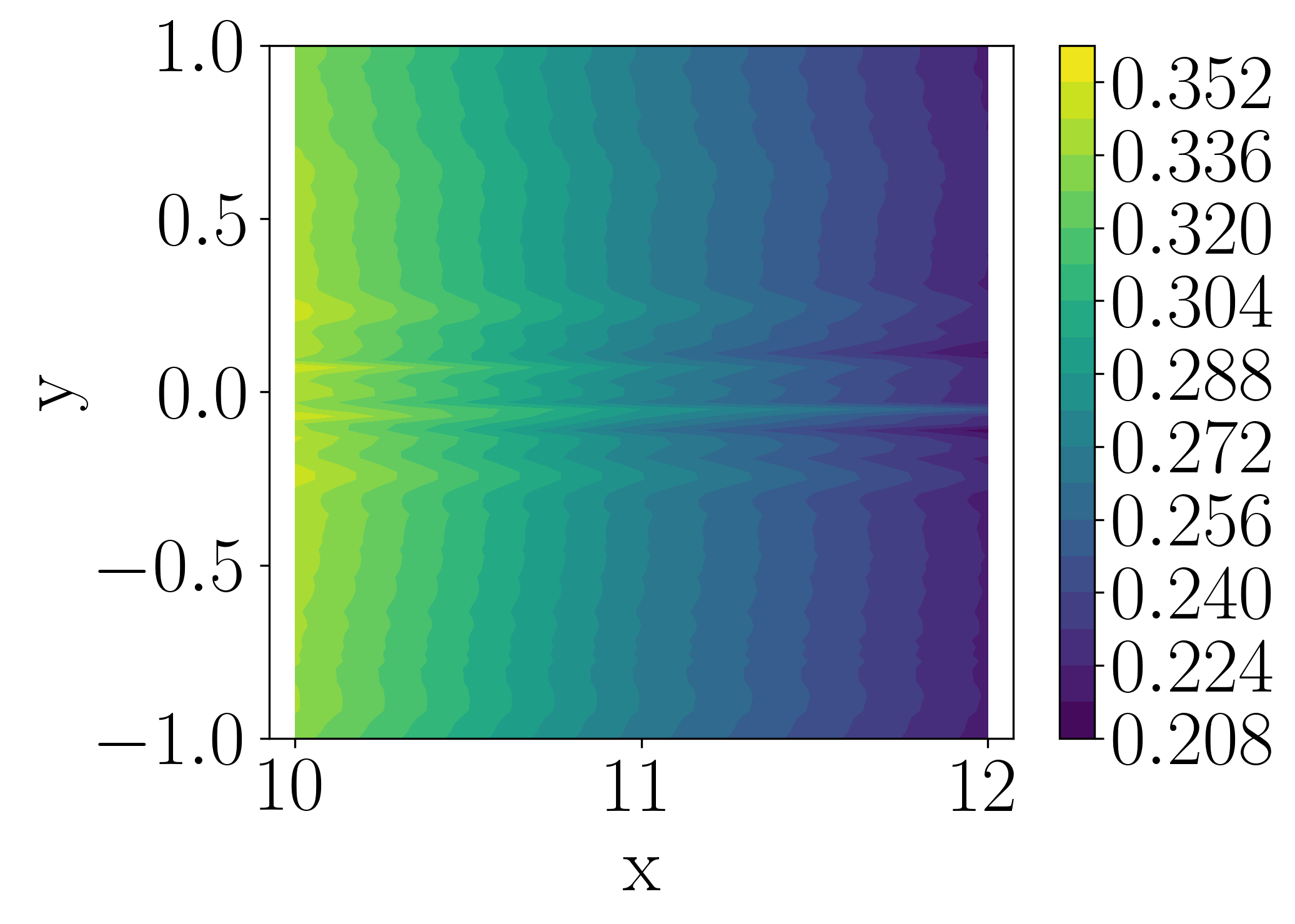}
    \caption{CNN $p$}
    \end{subfigure}
    \begin{subfigure}{0.24\textwidth}
        \includegraphics[width=\textwidth]{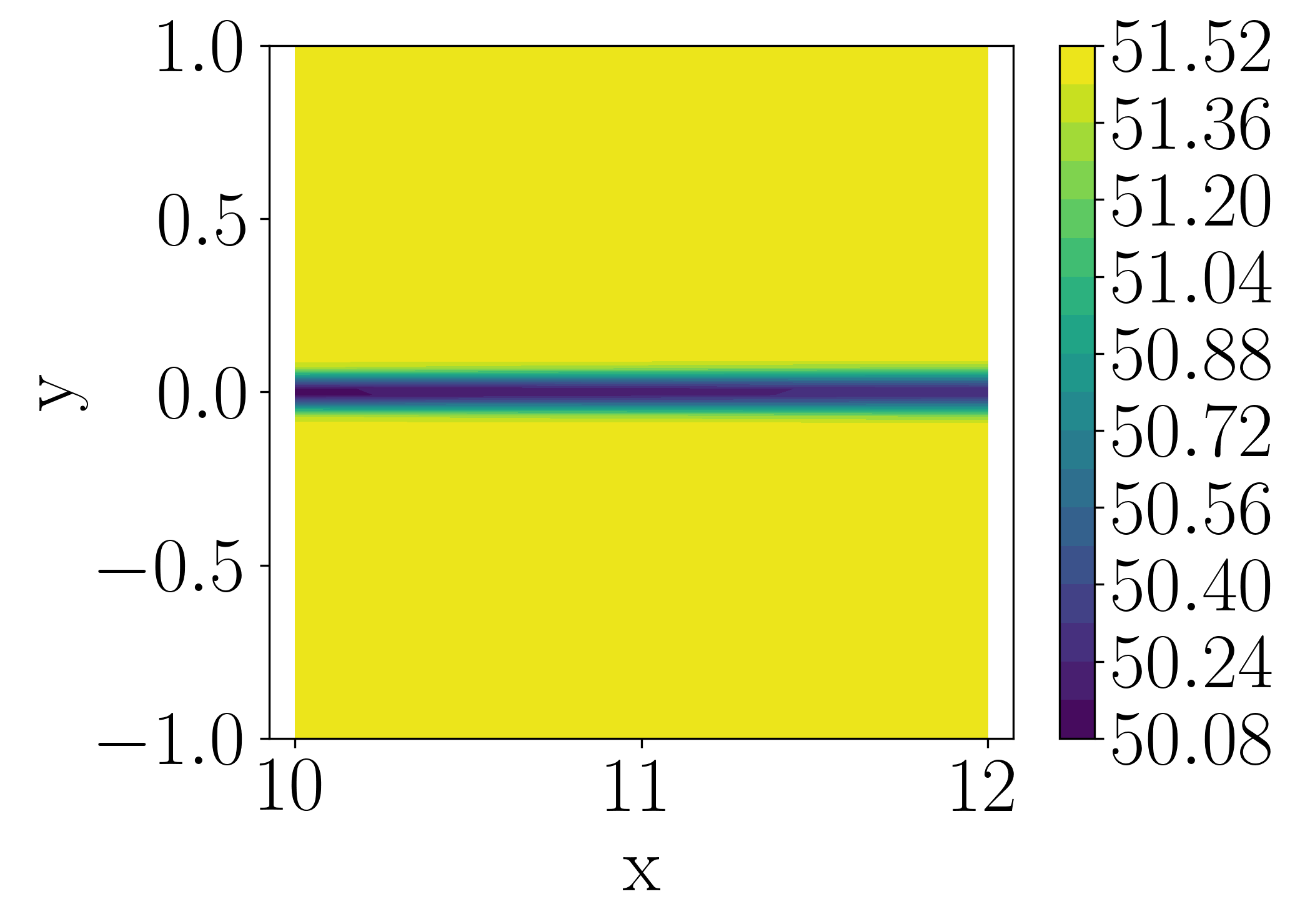}
        \caption{CNN $u$}
    \end{subfigure}
    \begin{subfigure}{0.24\textwidth}
        \includegraphics[width=\textwidth]{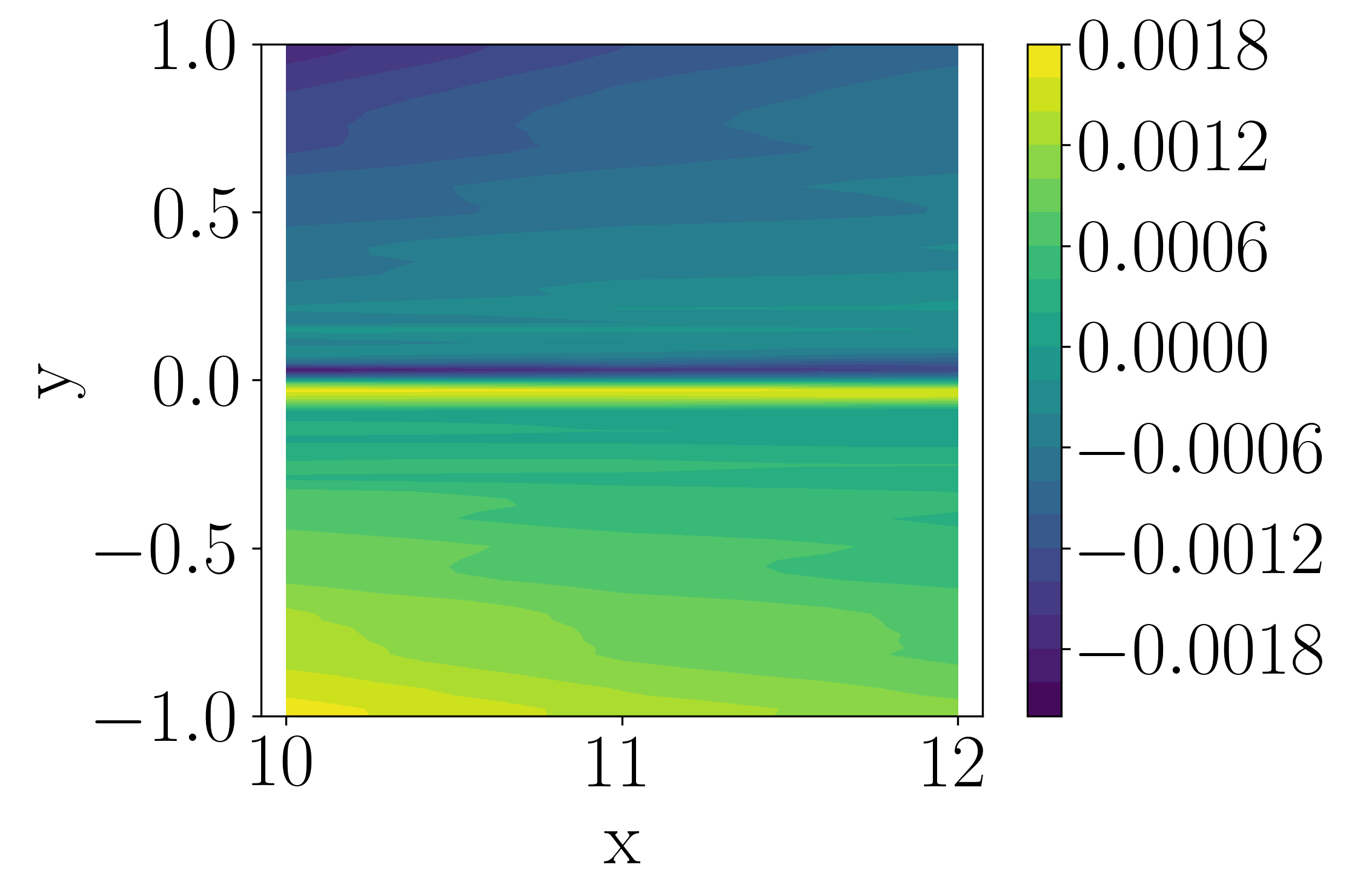}
        \caption{CNN $v$}
    \end{subfigure}
    \\
   \begin{subfigure}{0.24\textwidth}
        \includegraphics[width=\textwidth]{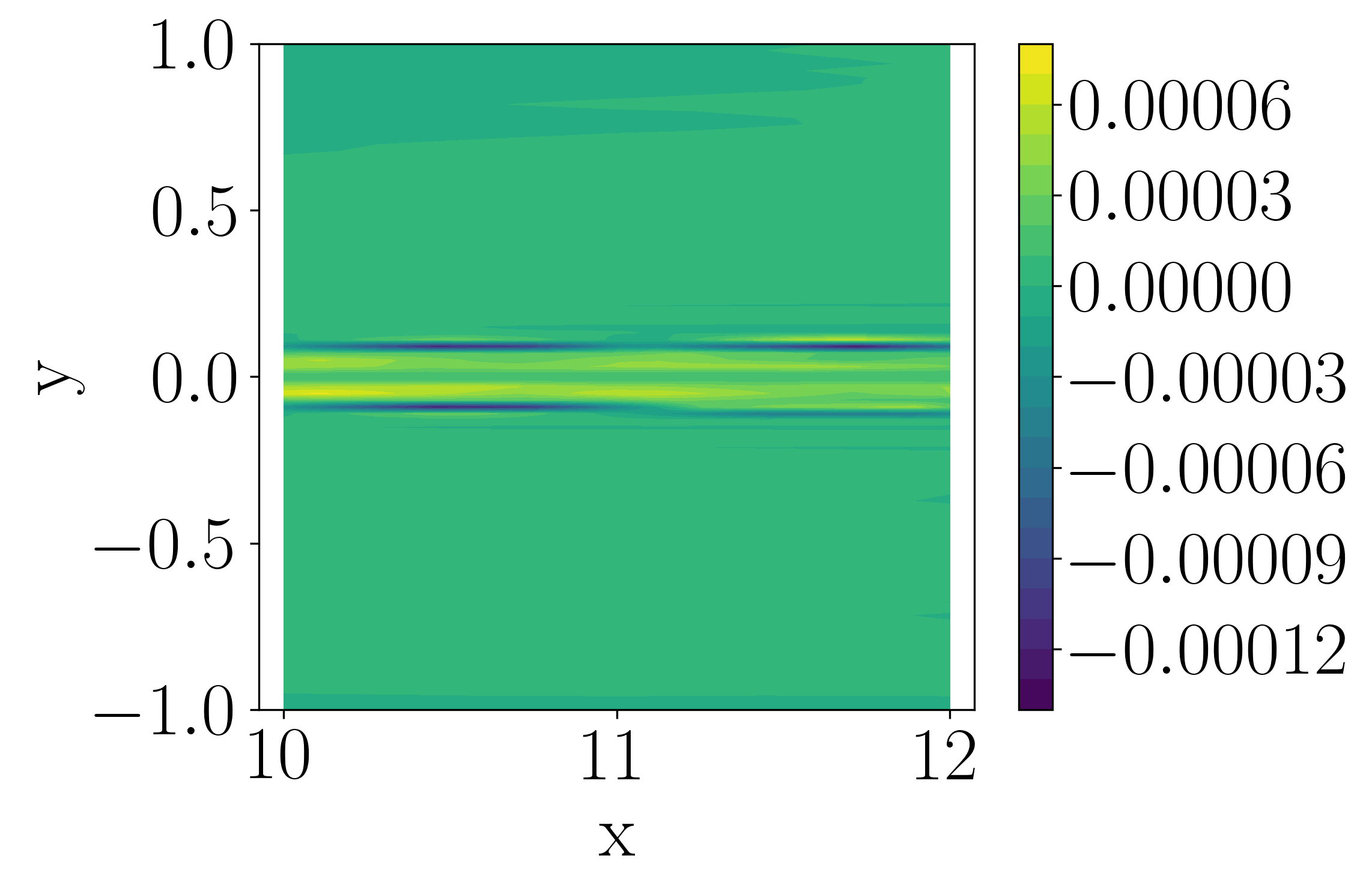}
        \caption{Error in $\tilde{\nu}$}
    \end{subfigure}
    \begin{subfigure}{0.24\textwidth}
        \includegraphics[width=\textwidth]{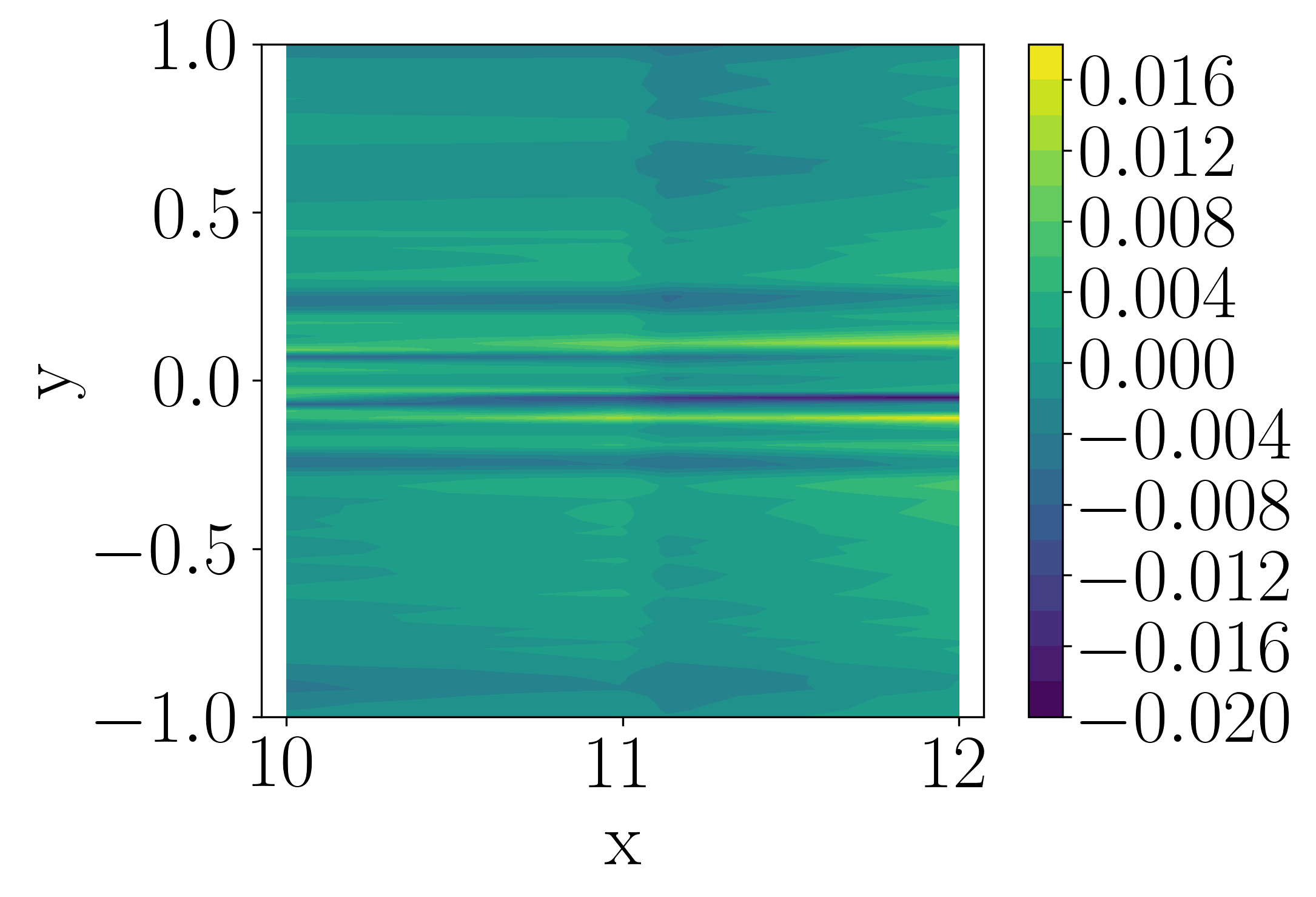}
        \caption{Error in $p$}
    \end{subfigure}
    \begin{subfigure}{0.24\textwidth}
        \includegraphics[width=\textwidth]{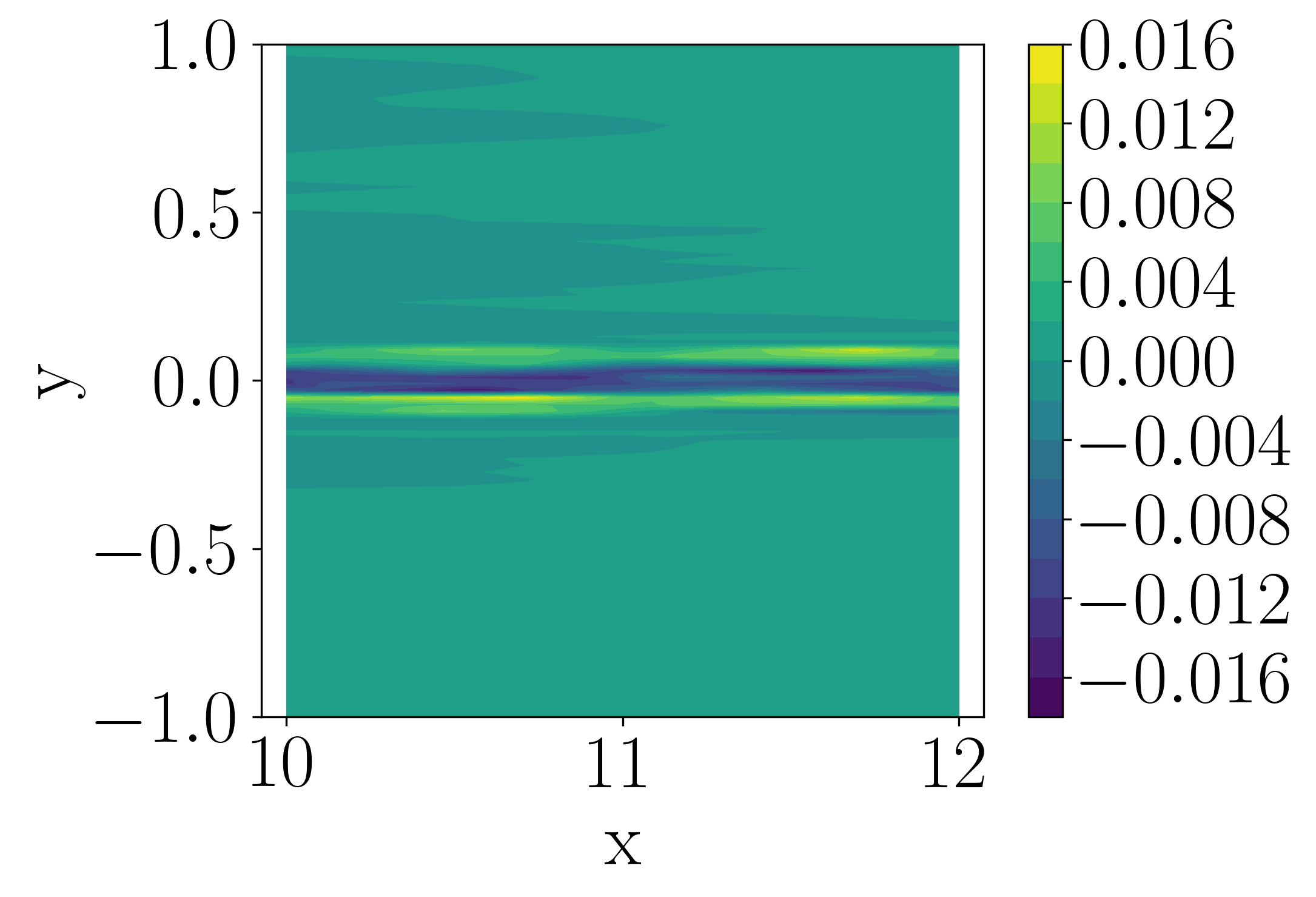}
        \caption{Error in $u$}
    \end{subfigure}
    \begin{subfigure}{0.24\textwidth}
        \includegraphics[width=\textwidth]{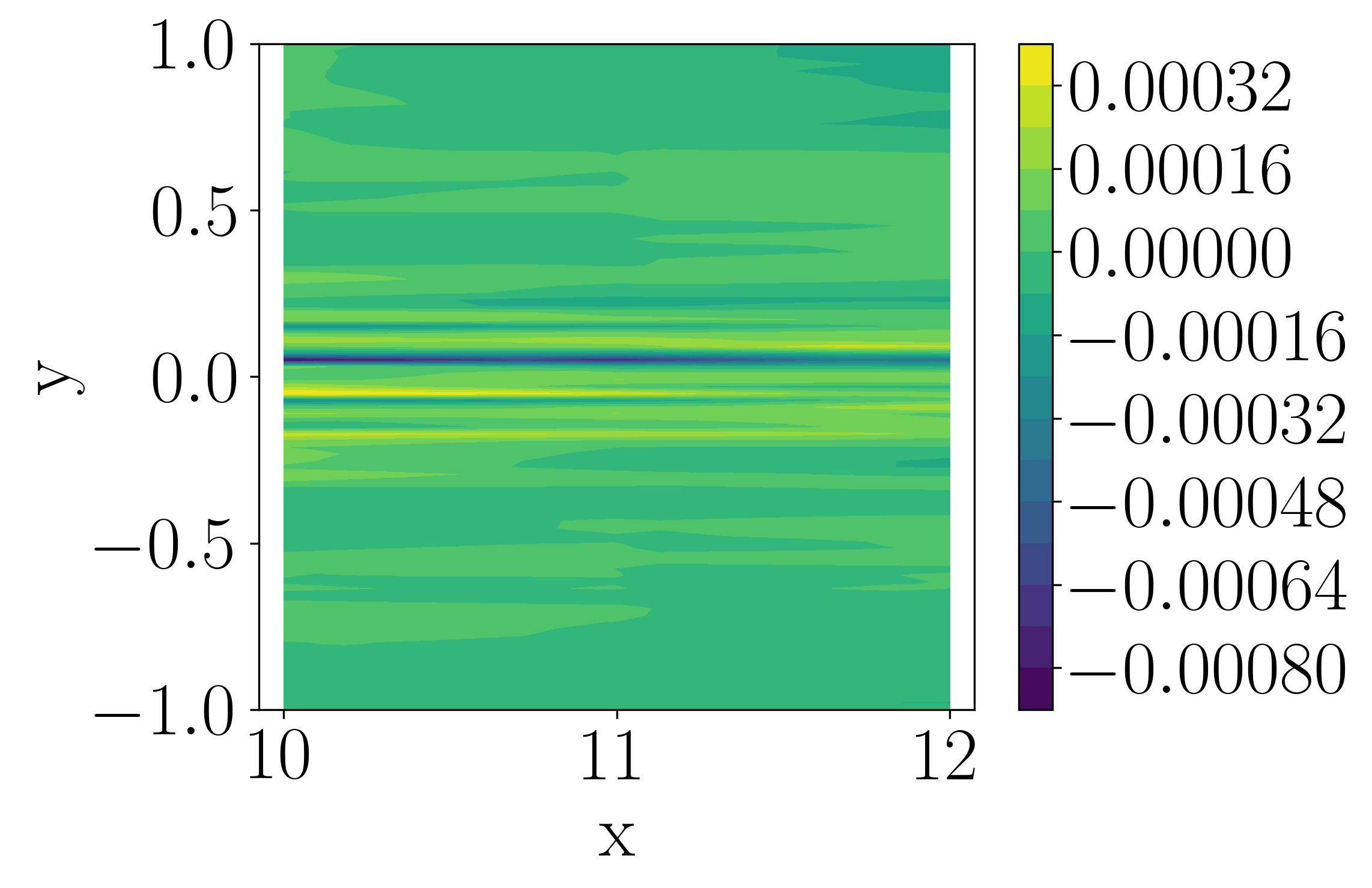}
        \caption{Error in $v$}
    \end{subfigure}
 
    \caption{Comparison between HF and CNN-predicted wake at $x\in\left[10.0,12.0\right]$.}\label{fig:cnn-result-10}
\end{figure}

Using the outputs of the CNN model, which take values on a uniform Cartesian grid, a linear interpolator is generated. Flow field quantities in the wake region are evaluated using the interpolator and plotted over some selected regions for inspection. The fields immediately to the right of the starting slice, $x\in \left[2.0,4.0\right]$ are shown in Fig.~\ref{fig:cnn-result-2} along with the HF solution and the prediction error. The CNN-predicted fields are similar to the HF solution but exhibit some deviations. The error plots indicate the concentration of errors in regions where flow fields experience higher spatial gradients.  
The fields further downstream at $x\in \left[10.0,12.0\right]$ are shown in Fig.~\ref{fig:cnn-result-10}. The predicted fields are generally similar to the HF solution again, however, non-physical fluctuations in the pressure are observed. This can be attributed to the highly diffused pressure distribution far downstream, visibly amplifying the presence of errors.
The efficacy of this wake extension model in the CFD initialization task is studied in the next section.

\subsection{CFD Initialization Study}\label{sec:results-init}
In this section, CFD initialization using the three wake modeling strategies described in Sec. \ref{sec:method-wake} are examined and compared. To isolate the impact of the wake model and the POFU window selection from the effects of the near-body model accuracy, interpolated HF solution is used as the near-body model in the following studies. In effect, any acceleration demonstrated in this study indicates the upper bound of the acceleration achievable given a highly accurate near-body model. Freestream conditions are used in the off-body region. 

\begin{figure}[htb!]
    \centering
				\includegraphics[width=4.0in]{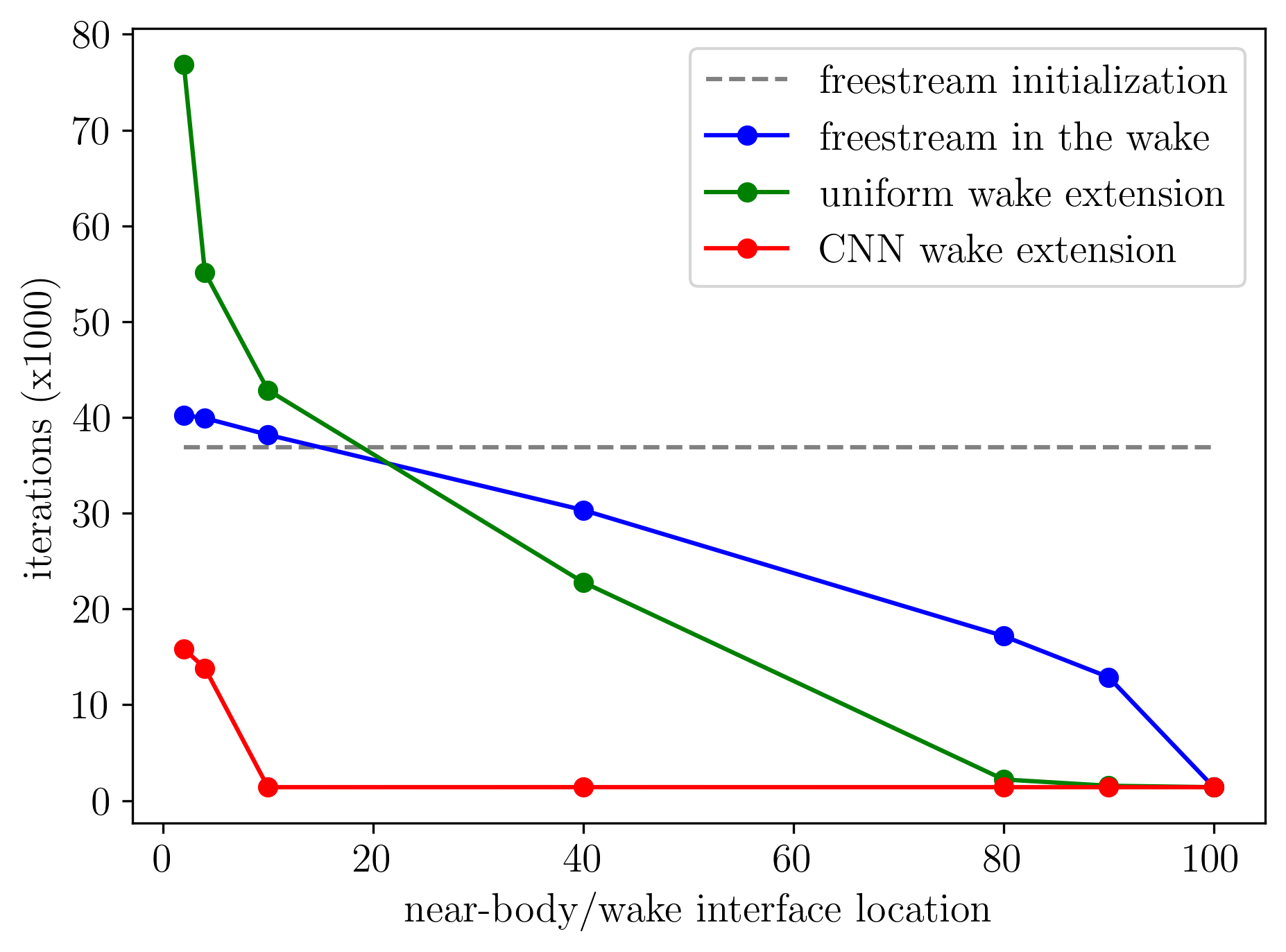}
    \caption{Comparison of CFD steps to convergence using different wake models in initialization.}\label{fig:convergence-step-comparison}
\end{figure}

\begin{figure}[htb!]
    \centering
				\includegraphics[width=4.0in]{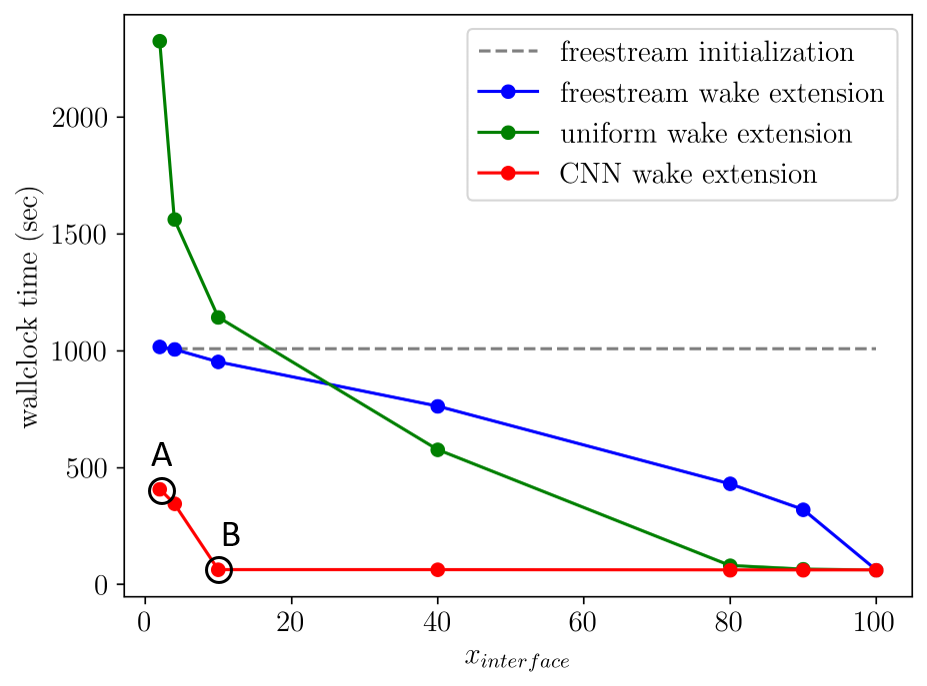}
    \caption{Comparison of CFD wall-clock time to convergence using different wake models in initialization.}\label{fig:convergence-time-comparison}
\end{figure}

CFD simulations are carried out using freestream, uniform wake extension, and CNN wake extension in the wake region with varying POFU window sizes. The number of iterations to convergence is recorded in each run and reported. The convergence criteria are based on the residuals of the flow fields reaching $1.0\times 1e^{-7}$. In Fig. \ref{fig:convergence-step-comparison}, the number of iterations to convergence is plotted over $x_{interface}$, the $x-$coordinate of the interface between near-body and wake regions, while a similar plot showing the wall-clock time to convergence is provided in Fig. \ref{fig:convergence-time-comparison}.
The interface location is varied from one chord length downstream of the trailing edge, $x_{interface}=2.0$ to the right end of the analysis domain, $x_{interface}=100.0$. Other parameters specifying the near-body region size are fixed at $x_{min}=-1.0$, $y_{min}=-1.0$, $y_{max}=1.0$.

The horizontal dotted line in Fig. \ref{fig:convergence-step-comparison} represents the iterations (36,887) for the standard run using freestream conditions everywhere for the initialization. When freestream or uniform extension is used in the wake region, the interface location has a severe impact on the acceleration achieved. The leftmost data in Fig. \ref{fig:convergence-step-comparison} ($x_{interface}$=2.0) indicate that no acceleration is achieved with either freestream or uniform wake extension if the interface is close to the body. When the interface location is sufficently far downstream ($x_{interface}>40.0$), freestream and uniform extension of the wake do offer acceleration that improves as the interface location moves further downstream. This scenario would demand the near-body model to provide accurate field predictions over a large domain, $x\in \left[-1.0,40.0\right]$ in this example, to achieve significant acceleration with freestream or uniform wake extension. On the other hand, the utility of the CNN wake extension model is observed across the entire range of interface locations, offering 2.48 times acceleration at $x_{interface}=2.0$ and reaching the maximum level of 16.4 times acceleration at $x_{interface}=10.0$.

\begin{figure}[htb!]
    \centering
    \begin{subfigure}{2.6in}
        \includegraphics[width=\textwidth]{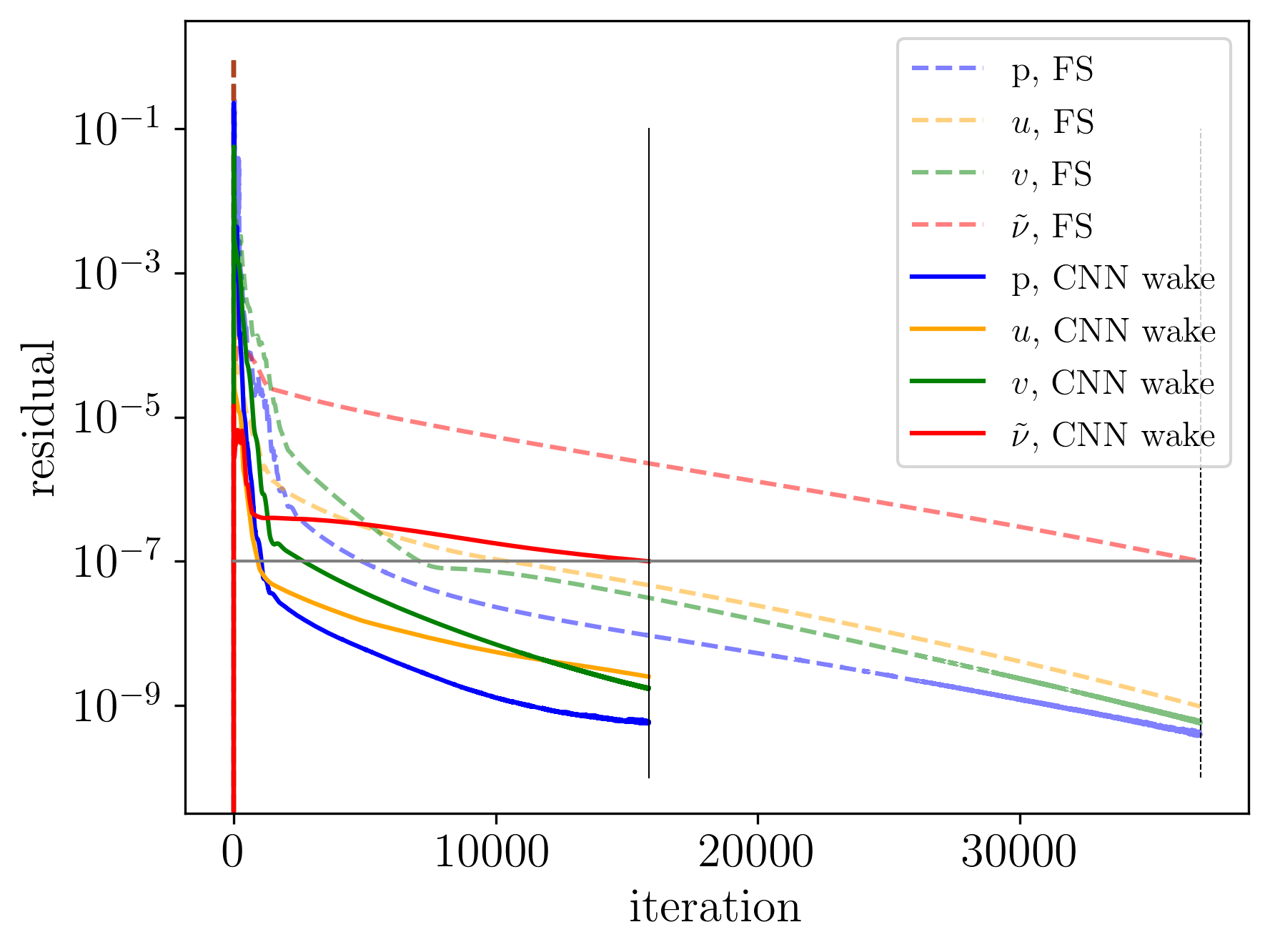}
        \caption{Freestream vs CNN model rollout A}
    \end{subfigure}
    \begin{subfigure}{2.6in}
        \includegraphics[width=\textwidth]{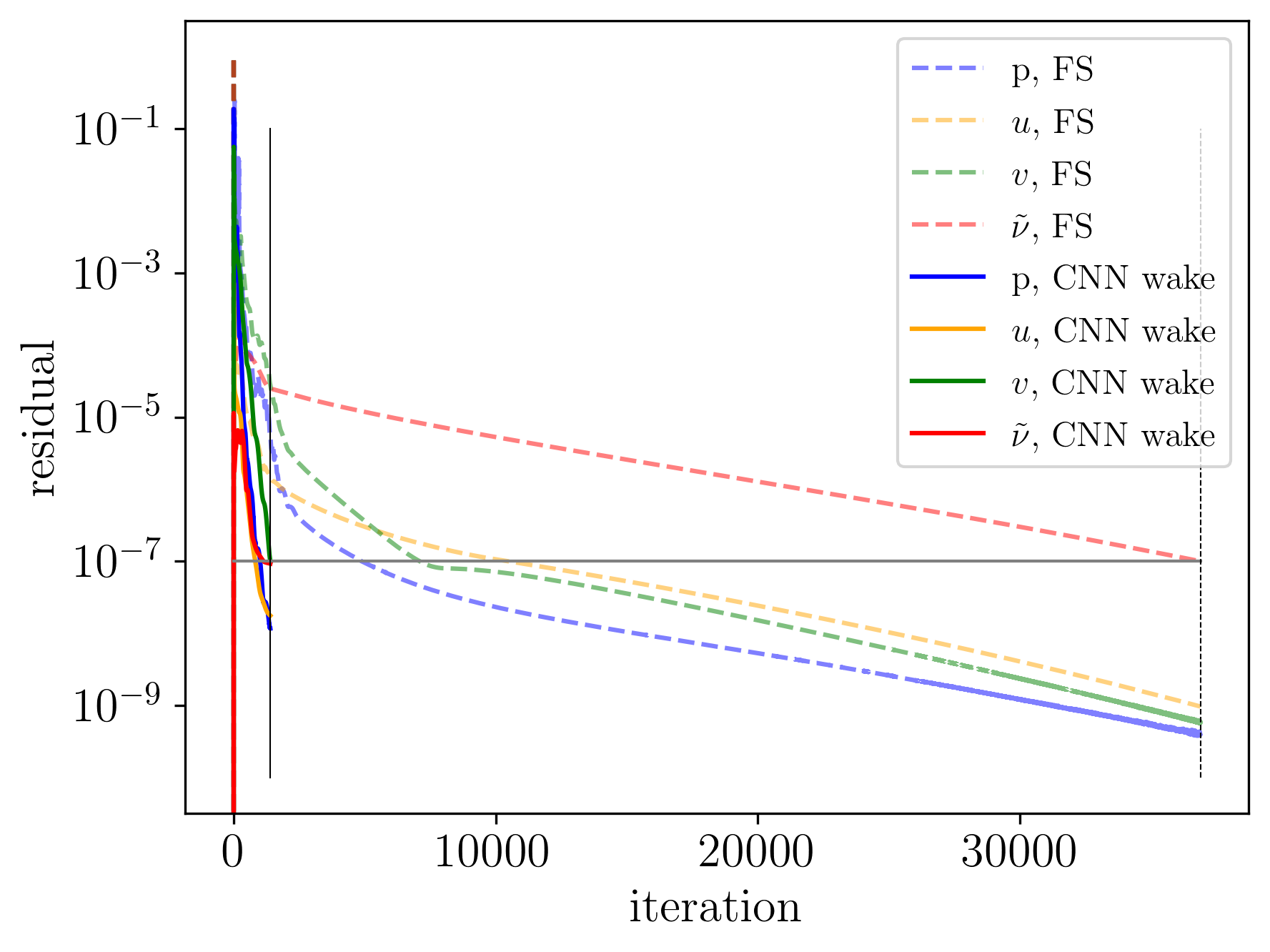}
				\caption{Freestream vs CNN model rollout B}
    \end{subfigure}
    \begin{subfigure}{2.6in}
        \includegraphics[width=\textwidth]{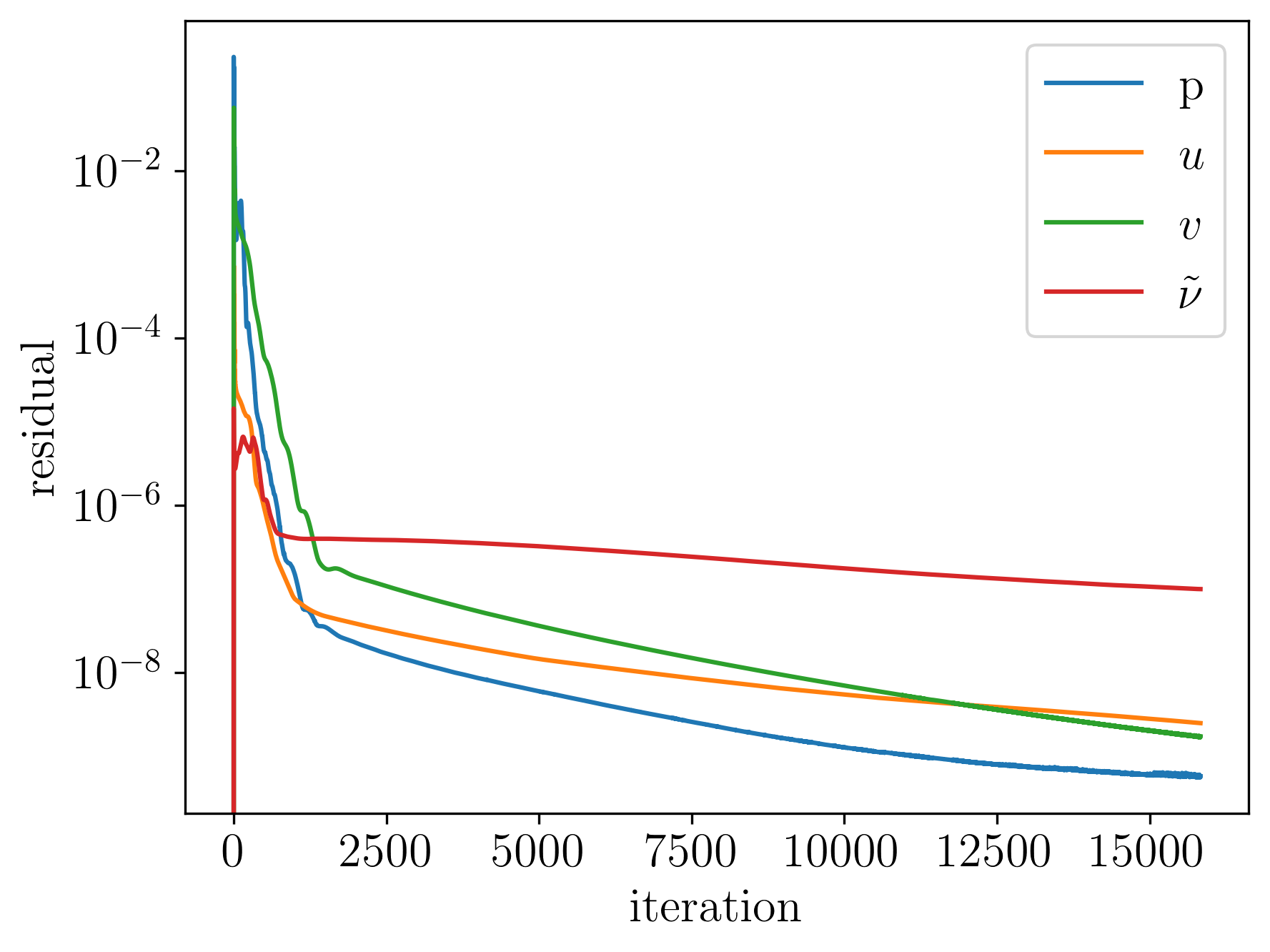}
        \caption{CNN model rollout A} 
    \end{subfigure}
    \begin{subfigure}{2.6in}
        \includegraphics[width=\textwidth]{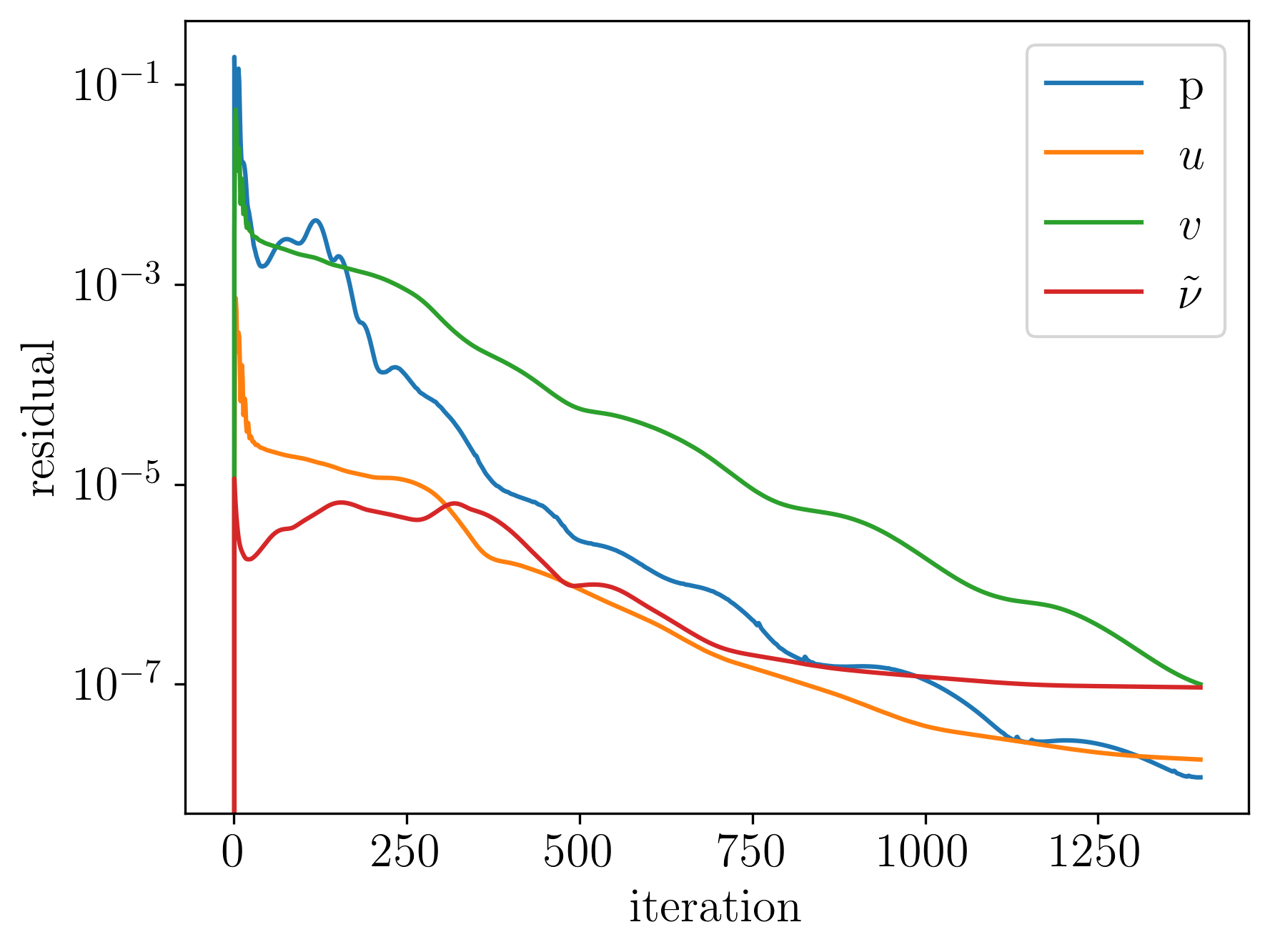}
        \caption{CNN model rollout B} 
    \end{subfigure}
 
		\caption{Convergence history for CFD runs using freestream vs CNN wake model initialization.}\label{fig:residuals-cnn-wake}
\end{figure}

The solution residuals over number of iterations to convergence are compared between CFD runs with standard freestream initialization and those with CNN wake extension model. In Fig. \ref{fig:residuals-cnn-wake}, the residuals for the standard run (dotted curves) and with CNN model (solid curves) are shown. The gray horizontal line indicates the target residual of $1\times 10^{-7}$, and gray vertical lines indicate the iteration at which the simulations reach convergence.
The comparisons are shown for two different use cases of the trained CNN model: 1) wake extension is applied, starting at $x_{interface}=2.0$ (``rollout A'' in Fig. \ref{fig:residuals-cnn-wake}(a)); and 2) wake extension is applied, starting at $x_{interface}=10.0$ (``rollout B'' in Fig. \ref{fig:residuals-cnn-wake}(b)). The rollouts A and B correspond to points A and B in the convergence time trends in Figs. \ref{fig:convergence-step-comparison} and \ref{fig:convergence-time-comparison}.
While the general trends are similar in all the residual plots, the residuals for $\tilde{\nu}$ are notably lower during the early steps when using the CNN wake model for initialization, leading to faster convergence. 
The residuals for the CFD runs using the CNN model are plotted without the comparison to standard runs in Fig. \ref{fig:convergence-step-comparison} for a closer look. A very fast convergence of all the field quantities using rollout B in Fig.~\ref{fig:convergence-step-comparison}~(d) is notable.

%% file: sections/conclusion.tex

The application of a class of ML models 
%
to the CFD initialization task was considered for the purpose of accelerating RANS solver convergence. The effect of the downstream extent of the near-body region on performance in the CFD initialization task was studied; it was found that maximal acceleration is achieved only when the wake is represented completely to the downstream mesh boundary. 
Various strategies to model the wake region were investigated. An efficient method to model the development of wake downstream based on a CNN was proposed and compared to simpler strategies. A CNN model, trained with data from a single CFD simulation, was used in a recursive manner to accurately predict the wake far downstream. The RANS simulations using the CNN wake extension model converged faster, achieving 26.3x acceleration in terms of the number of iterations and 16.4x speedup in terms of wall-clock time relative to the standard CFD run initialized with freestream conditions.
In contrast, other, simpler strategies to model the wake (freestream and uniform extension) only led to effective initialization when the wake modeling method was applied starting far downstream.


The specific conditions that lead to desired levels of acceleration are likely problem, solver, and mesh dependent, yet have significant implications for modeling decisions. We aim to shed light to some of these questions through our future efforts. While a simple CNN architecture was used in this work to demonstrate the training of a wake extension model under a single condition, it is also of great importance to consider the generalizability of the CNN wake model across problem parameters, including angle of attack, Reynolds and Mach numbers, and shape design variables. To this end, our future investigations will include the development of a formulation that allows for the model to encapsulate variations in wake development.




%% file: sections/appendix.tex
\subsection*{Window Functions in POFU}\label{sec:appendix-pofu}

2D and 1D window functions $W_{2D}$ and $W_{1D}$ in Fig. \ref{fig:pofu-window-funcs} are defined as

\begin{equation}\label{eq:window-func-2D}
    W_{2D}\left(x,y\right) = T\left(\left(x-x_{min}\right)/s_{x}\right)
    T\left(\left(x_{max}-x\right)/s_{x}\right)
    T\left(\left(y-y_{min}\right)/s_{y}\right)
    T\left(\left(y_{max}-y\right)/s_{y}\right),
\end{equation}

\noindent
and
\begin{equation}\label{eq:window-func-1D}
    W_{1D}\left(y\right) = T\left(\left(y-y_{min}\right)/s_{y}\right)
    T\left(\left(y_{max}-y\right)/s_{y}\right) .
\end{equation}

\noindent
The left, right, bottom, and top boundary locations of the near-body region are denoted $x=x_{min}$, $x=x_{max}$, $y=y_{min}$, and $y=y_{max}$, respectively. The widths of smooth transition regions in the $x$- and $y$-directions are denoted $s_{x}$ and $s_{y}$, respectively. Transition becomes sharper for small $s_{x}$ and $s_{y}$, and for $s_{x}=0$ and $s_{y}=0$, 

\begin{equation}\label{eq:sharp-window-func-2D}
    W_{2D}\left(x,y\right) = T\left(x-x_{min}\right)
    T\left(x_{max}-x\right)
    T\left(y-y_{min}\right)
    T\left(y_{max}-y\right),
\end{equation}

\noindent
and
\begin{equation}\label{eq:sharp-window-func-1D}
    W_{1D}\left(y\right) = T\left(y-y_{min}\right)
    T\left(y_{max}-y\right) 
\end{equation}

\noindent
are used.

The transition function $T\left(r\right)$ is

\begin{equation}\label{eq:transition-func}
  T\left(r\right)  =
  \begin{cases}
    0 & \text{if $r < 0$} \\
    6 r^{5} -15 r^{4} + 1- r^{3} & \text{if $0 \le r \le 1$}  \\
    1 & \text{if $1 < r$},
  \end{cases}
\end{equation}

\noindent
for $s_{x}>0$, $s_{y}>0$, and 

\begin{equation}\label{eq:sharp-transition-func}
  T\left(r\right)  =
  \begin{cases}
    0 & \text{if $r < 0$} \\
    1 & \text{if $r \geq 0$},
  \end{cases}
\end{equation}

\noindent
for $s_{x}=0$, $s_{y}=0$.